\documentclass[titlepage,12pt]{utarticle}
\usepackage{epsfig}
\usepackage{amsmath,amsfonts,latexsym}
\usepackage{array}
\usepackage[nosort]{cite}
\long\def\omit#1{}

\numberwithin{equation}{section}

\newcommand\CS{\mathcal{S}}
\newcommand\CV{\mathcal{V}}

\begin{document}

\preprint{UTTG--02--99\\
{\tt hep-th/9902153}\\
}

\title{Correlation Functions of Operators and Wilson Surfaces in the
$d=6,~(0,2)$ Theory in the Large $N$ Limit} 

\author{Richard Corrado, Bogdan Florea, 
	and Robert McNees\thanks{Research supported in part by
                the Robert A. Welch Foundation and NSF Grant
                PHY~9511632. \hfill\vskip .05em 
	\hspace*{.37em} 
	Email: {\tt rcorrado, bogdan, mcnees@zippy.ph.utexas.edu}}
}
	
\oneaddress{Theory Group, Department of Physics\\
        University of Texas at Austin\\
        Austin TX 78712 USA  }

\date{February 22, 1999}

\Abstract{We compute the two and three-point correlation functions of
chiral primary operators in the large $N$ limit of the $(0,2)$,
$d=6$ superconformal theory. We also consider the operator product
expansion of Wilson surfaces in the $(0,2)$ theory and compute the OPE
coefficients of the chiral primary operators at large $N$ from the
correlation functions of surfaces.
}

\maketitle

%%%%%%%%%%%%%%%%%%%%%%%%%%%%%%%%%%%%%%%%%%%%%%%%%%%%%%%%%%%%%%%%%%%%%

%%%
\renewcommand{\baselinestretch}{1.25} \normalsize

\section{Introduction}

The existence of the $(0,2)$ tensor multiplet in $d=6$ has been known for a
long time, but, until the last few years, there was no evidence suggesting 
the existence of {\it nontrivial} field theories for $d>4$.
This situation has changed with the discovery
that interacting $(0,2)$ superconformal theories exist in
compactifications of the IIB string on a K3
surface~\cite{Witten:Strings95}, as well as on the worldvolume of the
M-theory 5-brane~\cite{Strominger:Open-p-branes,Witten:IIBdual}. 
These theories are of great interest in themselves, but they are also
important because they have
applications~\cite{Rozali:Uduality,Berkooz-Rozali:FiveBrane,% 
Seiberg:NewTheories} to matrix theory~\cite{BFSS:Conjecture}, as well
to the study of 4-dimensional field theories, including the large $N$
limit of QCD~\cite{Witten:BranesQCD}.

Although many features of these nontrivial superconformal
field theories were evident within their string and M-theory
realizations, it is only recently that methods were developed to permit
detailed  computations within them. 
Important steps toward a deeper understanding of these
theories were made after 
realizing their importance in the compactification of matrix
theory. Soon afterwards, it was realized that
the $(0,2)$ theory had its own description as a matrix
model~\cite{Dijkgraaf:5D-BH-MatStrings,Aharony:Matrix-Six,%
Witten:NewGauge}.
This idea was carefully developed in~\cite{Aharony:02SCFT}, where the 
construction of states and the computation of correlation 
functions in a discrete light-cone quantization was discussed.

Another route towards the study of interacting superconformal field
theories appears in the correspondence between anti--de Sitter (AdS)
spaces and conformal field theories 
(CFTs)~\cite{Maldacena:LargeN,Gubser:Correlators,Witten:Holography}.
One aspect of this connection is that the large $N$
limit of the $(0,2)$ theory is described by supergravity on 
AdS$_7\times S^4$. The correspondence between SCFT operators and
supergravity Kaluza-Klein modes in this case was carried out in a
series of 
papers~\cite{Aharony:AdS,Minwalla:PrimaryOps,Leigh:Large-N,Halyo:AdS47}. 

In this paper, motivated by the success in computing correlation
functions of the $\CN=4$, $D=4$ SCFT from supergravity, we compute
correlation functions in the large $N$ limit of the $(0,2)$
theory. 

In section~\ref{sec:review}, we review several features of the $(0,2)$
theory, in particular the chiral primary operators that we will be
interested in for our computations. In
section~\ref{sec:correlators}, we perform the supergravity
computations that allow us to extract the 3-point function of the
chiral primary operators of the $(0,2)$ theory.
We use the quadratic corrections to the equations of motion to find
the necessary cubic terms in the 
action to  compute the 3-point function of the chiral primary
operators, as well as the 3-point function of two chiral primaries
with a certain higher dimensional operator. 
In an appendix, we point out some subtleties in the computation of the
quadratic action for the scalars corresponding to the chiral primary
operators.

In section~\ref{sec:surfaces}, we consider the operator product
expansion (OPE) of Wilson surfaces of the $(0,2)$ theory. We identify
the operators which are allowed to appear in the OPE and explicitly
compute the OPE coefficients of the chiral primary operators by
computing correlation functions of Wilson surfaces.

In the conclusion, we comment on our results and point to avenues of
further research. Several appendices collect results which were useful
in obtaining the correlation functions.

\section{Review of the Properties of the $(0,2)$ Theory}
\label{sec:review}

In the following, we will briefly review some of the features of the
$(0,2)$ field theory. For an additional discussion,
see~\cite{Seiberg:Sixteen} and the references therein. 

In six dimensions, the vector of $SO(5,1)$ appears in the
antisymmetric product of two $\mathbf{4}$s of the 
$Spin(4)\sim SU(2)\times SU(2)$ little group, 
$\mathbf{4}\times\mathbf{4} = \mathbf{6}_a\oplus \mathbf{10}_s$, so
that spinors in six dimensions may be taken to be
symplectic-Majorana-Weyl. The supercharges of $(0,2)$ SUSY,
$Q^a_\alpha$, carry a $Spin(4)$ spinor index, $\alpha$, and an
$Sp(2)_R\sim SO(5)$ index, $a$, labeling the $\mathbf{4}$. The
supersymmetry algebra is 
\begin{equation}
\{ Q^a_\alpha, Q^b_\beta \} = 
2 \omega^{ab} \gamma^\mu_{\alpha\beta} P_\mu 
+  \gamma^\mu_{\alpha\beta} Z^{ab}_\mu, \label{eq:zerotwoalg}
\end{equation}
where $\omega^{ab}$ is the $Sp(2)$-invariant tensor. A central charge
$Z^{ab}_\mu$ is allowed. This central charge must transform as a
Lorentz vector (as such, it couples to a string) and under
the  $\mathbf{5}$ of $Sp(2)_R$. The only massless representation
of~\eqref{eq:zerotwoalg} that does not involve a graviton is the tensor
multiplet. Labeling fields by their $SU(2)\times SU(2)\times Sp(2)_R$
representations, the tensor multiplet contains a 2-form, $B_{\mu\nu}$,
in the $(1,3;\mathbf{1})$ whose field strength $H=dB$ is (anti)
self-dual, four fermions, $\lambda^a_\alpha$, in the
$(2(1,2);\mathbf{4})$, and five scalars, $\phi^{ab}$, 
in the $(1,1;\mathbf{5})$.

The tensor multiplet appears on the M5-brane in the following
fashion. The M5-brane is a BPS state in $D=11$ supergravity which
preserves 16 of the 32 bulk supercharges. By considering the spectrum
of fluctuations around the 5-brane supergravity
solution, one finds 16 fermionic
and eight bosonic zero modes~\cite{Kaplan-Michelson:ZeroModes}. Five
of the bosonic zero modes are 
translational, corresponding to fluctuations in the directions
transverse to the brane, so they form a $\mathbf{5}$ of the $Spin(5)$
component of the broken 11-dimensional Lorentz group. The other 3
bosonic modes come from fluctuations of the bulk 3-form and are
arranged in an antisymmetric 2-form on the 5-brane worldvolume. The
fermionic degrees of freedom form a $\mathbf{4}$ of $Spin(5)$. This is
exactly the content of the $(0,2)$ tensor multiplet. An interacting
field theory is obtained when one considers a configuration of $N$
parallel 5-branes. In this case, M2-branes are allowed to stretch
between the 5-branes~\cite{Strominger:Open-p-branes}, leading to 
$N+1$ massless tensor multiplets on the worldvolumes of the
M5-branes. Since the worldvolume theory is chiral, there are no
additional massless states arising when the 5-branes are
coincident. The scalars in one linear combination of these tensor
multiplets corresponds to the center-of-mass motion of the collection
of 5-branes, so that tensor multiplet is free and decouples from the
rest of the theory. The moduli space of scalars in the remaining part
of the theory is $\BR^{5N}/S_{N+1}$. The scalars, $\phi^i_A$, are in
the irreducible $\mathbf{N}$ of $S_{N+1}$, labeled by $A$, while $i$
labels the $\mathbf{5}$ of the $SO(5)$ R-symmetry. The 
expectation values of the scalars determine the tensions of the BPS
strings which appear at the $(2\perp 5|1)$ brane intersections. Away
from the orbifold fixed points, the theory on the 5-branes is free. At
the fixed points of the $S_{N+1}$ group, some number of 5-branes
coincide. There the theory is an interacting superconformal field
theory.  

\subsection{Chiral Primary Operators of the $(0,2)$ Theory} 
\label{sec:cpos}

We are interested in the chiral primary operators of the $(0,2)$,
$d=6$, $U(N)$ field
theory.
In~\cite{Distler:Six-SCFT,Minwalla:UnitarityBounds,Aharony:02SCFT},
unitarity bounds on the dimensions of operators are obtained. 
For states which are Lorentz singlets and have $Sp(2)$ Dynkin labels
$(l,k)$, there are superconformal primaries at dimension $2(l+k)$ with
a null vector at level $\tfrac{1}{2}$, at dimension $2(l+k)+2$ with a
null vector at level $1$, at dimension $2(l+k)+4$ with a
null vector at level $\tfrac{3}{2}$, and at dimension $2(l+k)+6$ with
a null vector at level $2$. All of these states belong to short
multiplets, so we expect that the corresponding operators have
conformal dimensions which are protected. We will refer to the primary
operators of the shortest of these multiplets as chiral primary
operators. 
 
From these results, it is evident that a rank~$k$ symmetric,
traceless tensor representation of $SO(5)$, with $Sp(2)$ Dynkin label
$(0,k)$, is a chiral primary field of dimension $\Delta=2k$.
That such states actually exist in the $(0,2)$ theory was found in the
discrete light-cone quantization (DLCQ) approach
of~\cite{Aharony:02SCFT}. Away from the $S_{N+1}$ fixed point,
operators of these dimensions and quantum numbers have a realization
in terms of $S_{N+1}$ invariant, 
symmetric traceless (in the $SO(5)$ indices) products of $k$ free
scalars. At the fixed point, it is not clear that this realization 
makes any sense, so we will not pursue it further. We will simply
denote these chiral primaries as $\CO^I$.  

Correlation functions of these operators are restricted by
conformal invariance. The two-point function is determined up to a
free constant, which can be taken to be unity
\begin{equation}
\langle \CO^{I_1}(x_1) \CO^{I_2}(x_2) \rangle 
= \frac{\delta^{I_1 I_2}}{|x_1-x_2|^{4k}}.
\label{eq:const-2pt}
\end{equation}
Three-point functions are similarly constrained
\begin{equation}
\langle \CO^{I_1}(x_1) \CO^{I_2}(x_2) 
\CO^{I_3}(x_3) \rangle 
= \frac{c_3 \,
\langle C^{I_1}C^{I_2}C^{I_3}\rangle }{
|x_1-x_2|^{4\alpha_3} |x_2-x_3|^{4\alpha_1} |x_3-x_1|^{4\alpha_2}},
\label{eq:const-3pt}
\end{equation}
where we have denoted the 
symmetric, traceless $SO(5)$ tensors by $C^I_{i_1\cdots i_k}$, defined
such that the 
spherical harmonics on $S^4$ are 
$Y^I=C^I_{i_1\cdots i_k} x^{i_1}\cdots x^{i_k}$, with a  normalization
such that 
$C^{I_1}_{i_1\cdots i_k}C^{I_2i_1\cdots i_k}=\delta^{I_1 I_2}$. The
quantity 
\begin{equation}
\begin{gathered}
\langle C^{I_1}C^{I_2}C^{I_3}\rangle 
= C^{I_1}_{i_1\cdots i_{\alpha_2+\alpha_3}} 
{C^{I_2\, i_1\cdots i_{\alpha_3}}}_{j_1\cdots j_{\alpha_1}}
C^{I_3\,i_{\alpha_3+1}\cdots i_{\alpha_3+\alpha_2}j_1\cdots
j_{\alpha_1}}, \\
\alpha_i = \tfrac{1}{2}\sum_{j=1}^3 k_j - k_i ,\\
\Sigma = k_1 + k_2 +k_3,
\end{gathered}
\label{eq:3C-contract}
\end{equation}
is the $SO(5)$ invariant contraction of three $C^I_{i_1\cdots i_k}$.  
Our goal in section~\ref{sec:correlators} will be to obtain the
coefficient of this 3-point function in the large $N$ limit.

\subsection{The $(0,2)$, $d=6$ SCFT and Supergravity on AdS$_7\times S^4$} 

Information about the $(0,2)$ theory at the fixed point can be
obtained from its realization on the M5-brane worldvolume.
One begins by considering the M-Theory solution for a large number,
$N$, of coincident M5-branes. Next, the limit 
$M_P\rightarrow 0$ is taken, while keeping the tensions of the strings
on the 5-branes (due to membranes which stretch 
between the 5-branes) fixed. The result is M-Theory on AdS$_7\times
S^4$.  For large enough $N$, the curvature is very 
small in Planck units, so that classical supergravity is a good
approximation to the physics. On the other hand, this is 
precisely the same limit that leads to the decoupling of bulk modes
from the $(0,2)$ theory on the 5-brane worldvolume. One 
is lead to the conclusion that the large $N$ limit of the $(0,2)$ field theory is described by 11D supergravity on
AdS$_7\times S^4$~\cite{Maldacena:LargeN}.

A prescription for a generating functional for the correlation
functions of operators was given within the AdS/CFT correspondence
in~\cite{Gubser:Correlators,Witten:Holography}. Each operator in the
CFT which is chiral, so that its conformal dimension is protected
against renormalization, corresponds to a mode of the KK
compactification of the supergravity on the compact manifold. The mass
of the mode corresponding to an operator of dimension $\Delta$ is 
$m^2=\Delta(\Delta-d)$, where $d$ is the number of CFT spacetime
dimensions. Then correlation functions may be computed from the formula
\begin{equation}
\left\langle e^{\int \phi_0^I \CO^I} \right\rangle_{CFT}
\sim e^{-S_{\text{sugra}}[\phi_0^I]},  \label{eq:gen-funct}
\end{equation}
where $\phi_0$ is the boundary data for the supergravity mode $\phi^I$
corresponding to the operator $\CO^I$ and $S_{\text{sugra}}[\phi_0^I]$
is the effective action computed from all tree graphs with external
$\phi$ legs. 

In the case of the $(0,2)$ theory, this correspondence was discussed
by~\cite{Aharony:AdS,Minwalla:PrimaryOps,Leigh:Large-N,Halyo:AdS47}. 
The chiral primary operators have conformal weight $\Delta=2k$, so
they correspond to KK-scalars, $s^I$ with masses
$M_s^2=4k(k-3)$. These scalars arise as linear combinations of the
harmonic modes of the trace of the metric and the 3-form with indices
on the sphere. 

In order to compute 2 and 3-point functions in the large $N$ limit of
the $(0,2)$ theory via the
prescription~\eqref{eq:gen-funct}, the supergravity action must be
computed to 
cubic order in the modes $s^I$. In the next section, we discuss in
detail the identification of the mode $s^I$ and the other supergravity
modes which are related to it via constraint equations. We then follow
the methodology of~\cite{Lee:3point} to compute the cubic
action from quadratic corrections to the supergravity equations of
motion. The cubic action is then used to obtain 2 and 3-point
correlation functions of the $(0,2)$ theory at large $N$.

\section{Correlation Functions from Supergravity}
\label{sec:correlators}

The spectrum of states of supergravity on AdS$_7\times S^4$ have been
worked out via linearization around a Freund-Rubin background
in~\cite{Pilch:4-sphere,vanNieuwenhuizen:4-sphere}, as well as via the
oscillator formalism in~\cite{Gunaydin:s4-oscill}. Here we present the
equations of motion and background appropriate to the correspondence
with the $(0,2)$ theory and identify the modes $s^I$
corresponding to the CPOs. We then consider quadratic
corrections to the equation of motion~\cite{Lee:3point} and extract
the normalization of the quadratic 
action for $s^I$ and the cubic $s$-$s$-$s$ vertex. With this action,
we derive the large~$N$ result for the 3-point
function~\eqref{eq:const-3pt} of normalized operators. 

\subsection{11D Supergravity on AdS$_7\times S^4$: Background and
Scalar Fluctuations}

We start with the action for 11D supergravity 
\begin{equation}
\begin{split}
S_{\text{sugra}} = \frac{1}{2\kappa^2} \int d^{11}x \, 
& \left[ \sqrt{-G}  \left( R - \frac{1}{24} (F_{m_1\cdots m_4})^2
\right) \right. \\ 
& \hspace*{5mm} \left. 
+ \frac{2\sqrt{2}}{(144)^2} \epsilon^{m_1\cdots m_{11}} 
F_{m_1\cdots m_4} F_{m_5\cdots m_8} A_{m_9\cdots m_{11}} \right].
\end{split} \label{eq:11d-act}
\end{equation}
We choose units in which $R_{\text{AdS}}=1$,  then $R_{S_4} =1/2$ and 
\begin{equation}
\frac{1}{2\kappa^2}=\frac{2N^3}{\pi^5}.
\end{equation}
Varying~\eqref{eq:11d-act} with respect to the metric results in
the equation of motion (after eliminating $R$)
\begin{equation}
R_{mn}
=\frac{1}{6} \left( F_{m m_1\cdots m_3}{F_{n}}^{m_1\cdots m_3} 
- \frac{1}{12} g_{mn} F^2 \right). \label{eq:11d-eom}
\end{equation}
Varying with respect to the 3-form $A_{mnp}$ yields the Maxwell
equations 
\begin{equation}
\nabla^m F_{mm_1\cdots m_3} = - \frac{\sqrt{2}}{1152} 
{\epsilon_{m_1\cdots m_3}}^{m_4\cdots m_{11}} 
F_{m_4\cdots m_7} F_{m_8\cdots m_{11}}. \label{eq:maxwell}
\end{equation}

The AdS$_7\times S^4$ background with $R_{\text{AdS}}=1$ is given by
\begin{equation}
\begin{gathered}
ds^2 = \frac{1}{z^2} \left( dz^2 +\eta_{ij} dx^i dx^j \right) 
+ \frac{1}{4} d\Omega_4^2, ~~~~
\eta_{ij} = \text{diag}(-1,1,\ldots,1), \\
R_{\mu_1\cdots\mu_4}
= - (g_{\mu_1\mu_3}g_{\mu_2\mu_4}
-g_{\mu_1\mu_4}g_{\mu_2\mu_3} ) ,~~~
R_{\mu\nu}= -6 g_{\mu\nu} ,~~~R^{(7)}=-42, \\
R_{\alpha_1\cdots\alpha_4} 
= 4(g_{\alpha_1\alpha_3}g_{\alpha_2\alpha_4} 
-g_{\alpha_1\alpha_4}g_{\alpha_2\alpha_3} ),~~~
R_{\alpha\beta}= 12 g_{\alpha\beta} ,~~~~ R^{(4)}=48,\\
\bar{F}_{\alpha_1\cdots\alpha_4}
= 3\sqrt{2} \epsilon_{\alpha_1\cdots\alpha_4}.
\end{gathered} \label{eq:ads-background}
\end{equation}
where $\mu,\nu,\ldots$ correspond to AdS indices, while
$\alpha,\beta,\ldots$ correspond to the sphere. 
Now we would like to expand around the background ~\eqref{eq:ads-background}.
This is simplified if we Weyl-rescale the metric on AdS. Then the fluctuations are
defined as:
\begin{equation}
\begin{gathered}
G_{mn}=g_{mn}+h_{mn}, \\
h_{\mu\nu} = h^\prime_{(\mu\nu)}
+\left(\frac{h^\prime}{7} -\frac{h_2}{5}\right) g_{\mu\nu}, \\
h_{\alpha\beta}=h_{(\alpha\beta)} + \frac{h_2}{4}g_{\alpha\beta}, \\
F_{m_1\cdots m_4} = \bar{F}_{m_1\cdots m_4} + f_{m_1\cdots m_4},~~~ 
f_{m_1\cdots m_4} = 4 \nabla_{[m_1}a_{m_2m_3 m_4]},
\end{gathered} \label{eq:ads-fluctuations}
\end{equation}
where $(mn)$ denotes symmetrization and removal of the trace.
The dependence on the coordinates of the sphere is best determined by
expanding the fluctuations in a basis of spherical harmonics.  
With the gauge choices 
$\nabla^\alpha h_{(\alpha\beta)}=\nabla^\alpha h_{\mu\alpha}
=\nabla^\alpha a_{\alpha mn} = 0$, the fluctuations which we are
interested in have the expansions
\begin{equation}
\begin{gathered}
h^\prime_{(\mu\nu)}=\sum_I h^{\prime I}_{(\mu\nu)}Y^I ,~~~~~~
h_{(\alpha\beta)} = \sum_I \phi^I Y^I_{(\alpha\beta)} \\
h^\prime= \sum_I h^{\prime I}Y^I ,~~~~~~ h_2 = \sum_I h_2^I Y^I, \\
f_{\alpha_1\cdots\alpha_4}
= -6 \sum_I b^I \epsilon_{\alpha_1\cdots\alpha_4} 
\nabla^\beta\nabla_\beta Y^I, \\
f_{\mu\alpha_1\cdots\alpha_3}
= 6 \sum_I \nabla_\mu b^I \epsilon_{\alpha_1\cdots\alpha_4} 
\nabla^{\alpha_4} Y^I. 
\end{gathered}
\end{equation}

By expanding the $\alpha\beta$ components of~\eqref{eq:11d-eom} to linear
order and projecting onto 
linearly independent spherical harmonics, one finds the field
equations and constraint
\begin{equation}
-\frac{1}{8} \left[ 
\left( \nabla^\mu\nabla_\mu 
+ \frac{1}{10} \nabla^\gamma \nabla_\gamma
-72 \right) h_2^I 
+ \nabla^\gamma \nabla_\gamma h^{\prime I} 
- 192\sqrt{2} \nabla^\gamma \nabla_\gamma b^I \right] Y^I =0, 
\label{eq:rab-field1}
\end{equation}
\begin{equation}
-\frac{1}{2} \left[ \nabla^\mu\nabla_\mu  
+ \nabla^\gamma \nabla_\gamma - 8 \right] \phi^I Y^I_{(\alpha\beta)} =0,  
\label{eq:rab-field2}
\end{equation}
\begin{equation}
-\frac{1}{2} \left[  h^{\prime I} - \frac{9}{10} h_2^I \right] 
\nabla_{(\alpha}\nabla_{\beta)} Y^I=0.
\label{eq:rab-constraint}
\end{equation}
Among the $\mu\alpha$ components, we find the constraint 
\begin{equation}
\frac{1}{2} \left[ \nabla^\nu h^{\prime I}_{\mu\nu}
+ 3 \nabla_\mu \left( \frac{3}{40} h_2^I -\frac{1}{7} h^{\prime I} 
+6\sqrt{2} b^I \right) \right] \nabla_\alpha Y^I =0,
\label{eq:rmua-constraint}
\end{equation}
where we neglect terms involving $h_{\mu\alpha}$. From the
$\nabla^m F_{m\alpha_1\cdots\alpha_3}$ component of~\eqref{eq:maxwell},
we find another field equation,
\begin{equation}
-6 b^I \nabla_\alpha \nabla^\gamma\nabla_\gamma Y^I
-6 \left[\nabla^\mu\nabla_\mu   b^I 
- \frac{1}{2\sqrt{2}} h^{\prime I}
+\frac{3\sqrt{2}}{5} h_2^I \right] \nabla_\alpha Y^I =0. 
\label{eq:maxwell-field}
\end{equation}

When $\nabla_{(\alpha}\nabla_{\beta)} Y^I\neq0$,
\eqref{eq:rab-constraint} can be solved for $h^\prime$ in terms of
$h_2$. Then from~\eqref{eq:rab-field1} and~\eqref{eq:maxwell-field},
after applying the eigenvalue equation~\eqref{eq:sh-eigen} for the spherical
harmonics, we find the coupled equations of motion 
\begin{equation}
\left[ \nabla^\mu \nabla_\mu 
+ 4 \begin{pmatrix} 
-k(k+3) & 9 \\
-4 k(k+3) &  -k(k+3)-18 
\end{pmatrix} \right] 
\begin{pmatrix} 48\sqrt{2} b^I \\ h_2^I \end{pmatrix} Y^I = 0.
\label{eq:coupled-eom}
\end{equation} 
The mass eigenvalues and eigenvectors of this pair of
equations are
\begin{equation}
\begin{split}
s^I &= \frac{k}{2k+3} \left( h_2^I + 32\sqrt{2} (k+3) b^I \right),
~~~m_s^2 = 4k(k-3), ~~~k\geq 2, \\
t^I &= \frac{k+3}{2k+3} \left( h_2^I - 32\sqrt{2} k b^I \right)
, \hspace*{1.4cm} m_t^2 = 4 (k+3)(k+6),
~~~k\geq 0. 
\end{split} \label{eq:eigenvalue-s}
\end{equation}
The mode $s^I$ has the right mass to correspond to a $\Delta=2k$
operator. Since it has an expansion on $S^4$ in terms of scalar spherical
harmonics of degree $k\geq 2$, each mode transforms in the rank~$k$
symmetric traceless tensor representation of $SO(5)$. These are the
same representations as the chiral primaries~$\CO^I$, so we
conclude that the $s^I$ are the corresponding supergravity
modes~\cite{Aharony:AdS,Minwalla:PrimaryOps,Leigh:Large-N,Halyo:AdS47}. 

We will also be interested in the modes $\phi^I$ appearing
in~\eqref{eq:rab-field2}. Applying the eigenvalue formula for the
tensor spherical harmonics~\eqref{eq:tensor-eigen}, we find
from~\eqref{eq:rab-field2} that these have masses
\begin{equation}
m_\phi^2 = 4k(k+3), ~~k\geq 2. \label{eq:phi-mass}
\end{equation}
The $SO(5)$ representation of these modes is that of the rank~2
symmetric, traceless spherical harmonics, namely the $(2,k)$, where we
again use the $Sp(2)$ Dynkin labels to denote the representation. From
the bounds quoted in section~\ref{sec:cpos}, an operator in this
representation of dimension $\Delta=2(2+k)+2=2k+6$  will be
protected. This is precisely the conformal dimension one predicts from
the mass formula~\eqref{eq:phi-mass}.  We will denote these operators
as $\Phi^I$. 

\subsection{The Action to Cubic Order for the Modes $s^I$ and $\phi^I$}

Now we would like to derive the action to cubic order for these modes. 
We first identify the supergravity fields that depend on $s^I$. For
our purposes, we can ignore the mass eigenstate~$t^I$
in~\eqref{eq:eigenvalue-s}. This mode corresponds to a
scalar operator with dimension $\Delta=2k+12$ and is not of immediate
relevance to our discussion. 
We can therefore solve~\eqref{eq:eigenvalue-s} for $h_2^I$
and $b^I$ in terms of $s^I$. From the
constraint~\eqref{eq:rab-constraint}, the modes $h^{\prime I}$ are 
obtained in terms of $s^I$ through their dependence on
$h_2^I$. Additionally, we have to solve the
constraint~\eqref{eq:rmua-constraint} to find the dependence of
$h^{\prime I}_{(\mu\nu)}$ on $s^I$. To do this, we note that the only
way to get a symmetric 
traceless tensor from a scalar is to act with covariant derivatives,
so we choose the ansatz 
\begin{equation}
h^{\prime I}_{(\mu\nu)} = H^{\prime I}_{(\mu\nu)} 
+ \alpha  \nabla_{(\mu}\nabla_{\nu)} s^I, \label{eq:solve-constraint} 
\end{equation} 
where $H^{\prime I}_{(\mu\nu)}$ is independent of $s^I$ (so we can
ignore it for our purposes) and is divergenceless, 
$\nabla^\mu H^{\prime I}_{(\mu\nu)}=0$. We then 
compute the normalization $\alpha$ by putting $s^I$ on shell and
solving the constraint~\eqref{eq:rab-constraint}.  We find that
\begin{equation}
\begin{split}
h^I_{\alpha\beta} &= h^I_{(\alpha\beta)} 
+ g_{\alpha\beta} V(k) s^I, \\
h^I_{\mu\nu} &= W(k)\nabla_{(\mu}\nabla_{\nu)} s^I 
+ g_{\mu\nu} U(k) s^I, \\  
b^I &= X(k) s^I,
\end{split} \label{eq:s-dependence}
\end{equation}
where
\begin{equation}
\begin{gathered}
V(k)=\frac{1}{4},~~~ U(k)=-\frac{1}{14}, \\
W(k) = \frac{3}{16k(2k+1)},~~~X(k)=\frac{1}{32\sqrt{2}k}.
\end{gathered} \label{eq:s-dependence-coeffs}
\end{equation}

For the $\phi^I$ modes, from the supergravity
action~\eqref{eq:11d-act} we find the quadratic Lagrangian
\begin{equation}
\begin{gathered}
L[\phi^I] = -\frac{B_I}{2} \left[ (\nabla_\mu \phi^I)^2 + 4k(k+3)
(\phi^I)^2 \right],  \vspace*{2mm} \\
B_I = \frac{z(k)}{2} \, \frac{2N^3}{\pi^5}.
\end{gathered} \label{eq:phi-action}
\end{equation} 
We note that this action is canonical and that the mode dependence of
the normalization is due solely to the properties of the spherical
harmonics. In particular, no subtleties arising from the boundary of
AdS$_7$ are encountered.   

For the modes, $s^I$, the computation of the quadratic action is more
complicated. Imposing  the constraint~\eqref{eq:rmua-constraint} 
by simple substitution of the solution~\eqref{eq:solve-constraint}
leads to terms in the action with higher derivatives, of the form
$(\nabla_{(\mu}\nabla_{\nu)}s^I)^2$ and 
$(\nabla_\rho \nabla_{(\mu}\nabla_{\nu)}s^I)^2$. 
In~\cite{Lee:3point}, where the analogous modes on AdS$_5\times S^5$
were studied,  it was proposed to deal with these terms by computing
the action on shell as a function of boundary values. We find that
this procedure fails, both for AdS$_5\times S^5$ and AdS$_7\times
S^4$, to produce a normalization consistent with the quadratic
equations of motion. For the completeness of our discussion we present 
the details of these computations in 
Appendix~\ref{sec:quadaction} and comment on why they fail, but we
will not need the (misleading) 
results to compute the correlation functions.

It turns out that the simplest way of obtaining both the
quadratic\footnote{We would like to thank Sangmin Lee for pointing 
out to us that the computation of the $s$-$s$-$\phi$ vertex allows
the computation of the proper quadratic normalization for $s^I$.} and
cubic terms in the action for the $s^I$  is by computing
quadratic corrections to the equations of motion. Once we have these
corrections, we promote them to an action, 
as explained in~\cite{Lee:3point}. 
To this end, the quadratic corrections to
equations~\eqref{eq:rab-field1}, \eqref{eq:rab-constraint},
\eqref{eq:maxwell-field}, and~\eqref{eq:rab-field2}  are computed in 
appendix~\ref{sec:quadcorr}. 

With $Q_\phi^I$ from~\eqref{eq:Qphi}, we can compute the $s$-$s$-$\phi$
vertex. Note that the 
$ss$ corrections to~\eqref{eq:rab-field2} can be written in the form
\begin{equation}
\begin{split}
\left( \nabla^\mu\nabla_\mu - f_1  \right)
\phi^{I_1} = & 2  Q_\phi^{I_1} \\
= & D^\phi_{I_1 I_2 I_3} s^{I_2} s^{I_3}
+E^\phi_{I_1 I_2 I_3}  \nabla^\mu s^{I_2} \nabla_\mu s^{I_3}
+ F^\phi_{I_1 I_2 I_3}  \nabla^{(\mu} \nabla^{\nu)} s^{I_2} 
\nabla_{(\mu} \nabla_{\nu)} s^{I_3}. 
\end{split} \label{eq:phi-ss}
\end{equation}  
The change of variables
\begin{equation}
\begin{gathered}
\phi^{I_1} = \tilde{\phi}^{I_1} 
+ J^\phi_{I_1 I_2 I_3} \tilde{s}^{I_2} \tilde{s}^{I_3}
+L^\phi_{I_1 I_2 I_3} \nabla^\mu \tilde{s}^{I_2} 
\nabla_\mu\tilde{s}^{I_3}, \\
\vbox{\vskip5mm}
J^\phi_{I_1 I_2 I_3} = \frac{1}{2} E^\phi_{I_1 I_2 I_3} 
- \frac{1}{4} ( m_2^2 + m_3^2 -f_1 -12) F^\phi_{I_1 I_2 I_3}, \\
\vbox{\vskip5mm}
L^\phi_{I_1 I_2 I_3} = \frac{1}{2} F^\phi_{I_1 I_2 I_3} ,
\end{gathered}
\end{equation}
leaves the normalization of the  quadratic action for $s^I$ and
$\phi^I$ invariant, but removes the derivatives from the right-hand
side of~\eqref{eq:phi-ss}
\begin{equation}
\left( \nabla^\mu\nabla_\mu - f_1  \right)
\tilde \phi^{I_1} = 
\lambda^\phi_{I_1, I_2 I_3} \tilde{s}^{I_2} \tilde{s}^{I_3} ,
\end{equation}
where
\begin{equation}
\begin{split}
\lambda^\phi_{I_1, I_2 I_3} &=  D^\phi_{I_1 I_2 I_3}  
-  ( m_2^2 + m_3^2 -f_1)  J^\phi_{I_1 I_2 I_3}
- \frac{2}{7} m_2^2 m_3^2 L^\phi_{I_1 I_2 I_3}   \\
\vbox{\vskip1cm}
&= - \frac{9a(k_1,k_2,k_3)}{64z(k_1)} \, 
\frac{\alpha_1(\alpha_1-1)(2\alpha_1-3)
\Sigma(\Sigma+1)(\Sigma+3)}{
k_2 (2k_2+1) k_3 (2k_3+1)}  \, 
\langle T^{I_1} C^{I_2} C^{I_3} \rangle.
\end{split} 
\end{equation}  
The cubic $s$-$s$-$\phi$ vertex will be obtained from this by
multiplication by the normalization $B_I$ 
of the quadratic action~\eqref{eq:phi-action} for $\phi^I$,
\begin{equation}
G^{\phi ss}_{I_1, I_2 I_3} 
= -\frac{9 N^3}{64 \pi^5} \, \frac{ \alpha_1(\alpha_1-1)(2\alpha_1-3)
\Sigma(\Sigma+1)(\Sigma+3)}{
 k_2 (2k_2+1) k_3 (2k_3+1)} \,  a(k_1,k_2,k_3) \, 
\langle T^{I_1} C^{I_2} C^{I_3} \rangle. \label{eq:phiss-vertex}
\vbox{\vskip8mm}
\end{equation}
\vbox{\vskip1mm}

This vertex may also be computed by studying corrections of the form
$s\phi$ to the equation of motion~\eqref{eq:eom-s} for the $s^I$
modes, which take the form
\begin{equation}
\begin{split}
\left( \nabla^\mu\nabla_\mu - 4k_1(k_1-3)\right) s^{I_1} = & 
D^{(s\phi)}_{I_1 I_2 I_3} \phi^{I_2} s^{I_3}
+E^{(s\phi)}_{I_1 I_2 I_3}  \nabla^\mu \phi^{I_2} \nabla_\mu s^{I_3}
 \\
\vbox{\vskip5mm}
&
+ F^{(s\phi)}_{I_1 I_2 I_3}  \nabla^{(\mu} \nabla^{\nu)} \phi^{I_2} 
\nabla_{(\mu} \nabla_{\nu)} s^{I_3} .
\end{split}
\end{equation}
Again, by a change of variables, this time with
\begin{equation}
\begin{gathered}
s^{I_1} = \tilde{s}^{I_1} 
+ J^{(s\phi)}_{I_1 I_2 I_3} \tilde{\phi}^{I_2} \tilde{s}^{I_3}
+L^{(s\phi)}_{I_1 I_2 I_3} \nabla^\mu \tilde{\phi}^{I_2} 
\nabla_\mu\tilde{s}^{I_3},  \\
\vbox{\vskip8mm}
J^{(s\phi)}_{I_1 I_2 I_3} = \frac{1}{2} E^{(s\phi)}_{I_1 I_2 I_3} 
- \frac{1}{4} ( f_2+ m_3^2 - m_1^2  -12) F^{(s\phi)}_{I_1 I_2 I_3},
 \\
\vbox{\vskip8mm}
L^{(s\phi)}_{I_1 I_2 I_3} = \frac{1}{2} F^{(s\phi)}_{I_1 I_2 I_3} ,
\end{gathered}
\end{equation}
we can remove the higher derivative
terms, obtaining
\begin{equation}
\begin{gathered}
\left( \nabla^\mu\nabla_\mu - 4k_1(k_1-3)\right) \tilde{s}^{I_1}
=  \lambda^{(s\phi)}_{I_2, I_1 I_3} \tilde{\phi}^{I_2}
\tilde{s}^{I_3},  \vspace*{2mm}\\
\vbox{\vskip8mm}
\lambda^{(s\phi)}_{I_1, I_2 I_3} = 
-\frac{a(k_1,k_2,k_3)}{4 z(k_2)}\, 
\frac{ \alpha_1(\alpha_1-1)(2\alpha_1-3)
\Sigma(\Sigma+1)(\Sigma+3)}{
(k_2-1) (2k_2+3) k_3 (2k_3+1)}  \,
\langle T^{I_1} C^{I_2} C^{I_3} \rangle.
\end{gathered} \label{eq:sphi-s-eom}
\end{equation}

The $\phi$-$s$-$s$ vertex in~\eqref{eq:phiss-vertex} should be
obtained by 
multiplying $\lambda^{(s\phi)}_{I_1, I_2 I_3}$ by the quadratic
normalization $A_{I_2}$ for the mode $s^{I_2}$. By comparison of these
expressions, taking into account a factor of two because the vertex is 
quadratic in the $s^I$,  we find that the linearized equation of
motion for $s^I$ can be obtained from a canonical quadratic action
\begin{equation}
\begin{split}
L[s^I] = 
&- \sum_I \frac{A_I}{2} \bigl[ (\nabla s^I)^2  +4k(k-3)(s^I)^2 \bigr], \\
A_{I} =&   \frac{9(k-1)(2k+3) z(k)}{16k(2k+1)}
\, \frac{2N^3}{\pi^5}. \label{eq:s-norm-true}
\end{split}
\end{equation}

The $s$-$s$-$s$ vertex is computed by considering the $ss$ terms
in~\eqref{eq:eom-s} 
\begin{equation}
\begin{split}
\left( \nabla^\mu\nabla_\mu - 4k_1(k_1-3)\right) s^{I_1} = & 
D_{I_1 I_2 I_3} s^{I_2} s^{I_3}
+E_{I_1 I_2 I_3}  \nabla^\mu s^{I_2} \nabla_\mu s^{I_3} \\
\vbox{\vskip5mm}
&
+ F_{I_1 I_2 I_3}  \nabla^{(\mu} \nabla^{\nu)} s^{I_2} 
\nabla_{(\mu} \nabla_{\nu)} s^{I_3} ,
\end{split}
\end{equation}
where the higher derivative
terms are removed by the change of variables
\begin{equation}
\begin{gathered}
s^{I_1} = \tilde{s}^{I_1} 
+ J_{I_1 I_2 I_3} \tilde{s}^{I_2} \tilde{s}^{I_3}
+L_{I_1 I_2 I_3} \nabla^\mu \tilde{s}^{I_2} 
\nabla_\mu\tilde{s}^{I_3}, \\
\vbox{\vskip8mm}
J_{I_1 I_2 I_3} = \frac{1}{2} E_{I_1 I_2 I_3} 
- \frac{1}{4} ( m_2^2+ m_3^2 -m_1^2 -12) F_{I_1 I_2 I_3},  \\
\vbox{\vskip8mm}
L_{I_1 I_2 I_3} = \frac{1}{2} F_{I_1 I_2 I_3} ,
\end{gathered}
\end{equation}
so that
\begin{equation}
\begin{gathered}
\left( \nabla^\mu\nabla_\mu - 4k_1(k_1-3)\right) \tilde{s}^{I_1}
=  \lambda_{I_1 I_2 I_3} \tilde{s}^{I_2} \tilde{s}^{I_3}, \\
\vbox{\vskip8mm}
\lambda_{I_1 I_2 I_3} = 
-\frac{3a(k_1,k_2,k_3)}{32 z(k_1)} \, 
\frac{ \alpha_1\alpha_2\alpha_3
(\Sigma-2)(\Sigma^2-1)(\Sigma^2-9)}{
(k_1-1) (2k_1+3) k_2 (2k_2+1)k_3 (2k_3+1)}  \,
\langle C^{I_1} C^{I_2} C^{I_3} \rangle. 
\end{gathered} \label{eq:lambda-s-s-s}
\end{equation}
Multiplying this by $A_{I_1}$, as given
in~\eqref{eq:s-norm-true}, we obtain  a totally symmetric function
\vspace*{1mm} 
\begin{equation}
\vbox{\vskip5mm}
G_{I_1 I_2 I_3} = 
-\frac{27N^3}{256\pi^5 } \,
\frac{ \alpha_1\alpha_2\alpha_3
(\Sigma-2)(\Sigma^2-1)(\Sigma^2-9)}{
 k_1 (2k_1+1) k_2 (2k_2+1)k_3 (2k_3+1)}  \,
a(k_1,k_2,k_3) \,
\langle C^{I_1} C^{I_2} C^{I_3} \rangle. \vspace*{2mm}
\end{equation}

Putting these results together, we have computed the following terms
in the action for the modes $s^I$ and $\phi^I$
\begin{equation}
\begin{split}
L[s^I,\phi^I] = 
&- \sum_I \left[ \frac{A_I}{2} \bigl[ (\nabla s^I)^2  +4k(k-3)(s^I)^2 \bigr]
+\frac{B_I}{2} \bigl[ (\nabla_\mu \phi^I)^2 + 4k(k+3)
(\phi^I)^2 \bigr] \right] \\
&+ \sum_{I_1,I_2,I_3} \left[ 
G^{\phi ss}_{I_1, I_2 I_3} \phi^{I_1} s^{I_2} s^{I_3}
+ \frac{1}{3} G_{I_1 I_2 I_3} s^{I_1} s^{I_2} s^{I_3} \right]
+ \cdots.
\end{split} \label{eq:quad-cubic-action}
\end{equation}

\subsection{Two and Three-Point Correlation Functions at Large $N$}

With an action for the modes~$s^I$ computed to cubic order, it is now
simple to compute correlation functions of the corresponding chiral
primary operators. We are assuming that the operators
are normalized~\eqref{eq:gen-funct}, so the coupling appearing in the
generating functional could involve a proportionality constant, 
{\it i.e.}, 
\begin{equation}
\left\langle e^{\int \CN^I s_0^I \CO^I} \right\rangle_{CFT}
\sim e^{-S_{\text{sugra}}[s_0^I]}. 
\end{equation}
The $\CN^I$ will be chosen such that the  large $N$ two-point function
is normalized. From the
formula~\eqref{eq:two-point-gen} that generates the 
two-point function, we find 
\begin{equation}
\langle \CO^{I_1}(x_1) \CO^{I_2}(x_2) \rangle = 
\frac{1}{(\CN^I)^2}\, \frac{1}{A_I}\, 
\frac{4(k-1)(2k-1)(2k-3)^2}{\pi^3}
\frac{\delta^{I_1I_2}}{
|\vec{x}_1-\vec{x}_2|^{4k}}, 
\label{largeN-2pt}
\end{equation} 
so that we set
\begin{equation}
\CN^I= -\frac{2^{3(k+1)/2}(2k-3)}{3N^{3/2}} \, 
\sqrt{\frac{(2k-1)(2k+1)\Gamma\bigl(k+\tfrac{3}{2}\bigr)}{\Gamma(k)}} .
\end{equation}
We have chosen a minus sign in this expression so that the coefficient
of the 3-point function below is positive.

From this normalization and the formula for the three-point
function~\eqref{eq:three-point-form}, we find the three-point function
for chiral primary operators, valid in the large $N$ limit,
\begin{equation}
\begin{split}
&\langle  \CO^{I_1}(x_1) \CO^{I_2}(x_2) \CO^{I_3} (x_3) \rangle \\
\vbox{\vskip1cm}
&=  
\frac{1}{
\sqrt{\pi}N^{3/2}} \, 
\left[ \prod_{i=1}^3 
\Gamma\bigl(\alpha_i+\tfrac{1}{2}\bigr)
\sqrt{\frac{4k_i^2-1}{
\Gamma(k_i)\Gamma\bigl(k_i+\tfrac{3}{2}\bigr)}} \, \right]  
\frac{2^{(\Sigma/2-15)/2} \Gamma\bigl(\tfrac{\Sigma}{2}\bigr)
\langle C^{I_1} C^{I_2} C^{I_3}\rangle }{
|\vec{x}_1-\vec{x}_2|^{4\alpha_3}
|\vec{x}_2-\vec{x}_3|^{4\alpha_1}
|\vec{x}_3-\vec{x}_1|^{4\alpha_2}}  \vspace*{5cm} 
\end{split} \label{eq:three-point-function}
\end{equation}
In terms of the conformal weights, we can express
this as
\begin{equation}
\begin{split}
&\langle  \CO^{I_1}(x_1) \CO^{I_2}(x_2) \CO^{I_3} (x_3) \rangle \\
\vbox{\vskip1cm}
& = 
\frac{2^{\tfrac{\Delta_1+\Delta_2+\Delta_3-30}{4}} 
\Gamma\bigl(\tfrac{\Delta_1+\Delta_2+\Delta_3}{2}\bigr) }{
\sqrt{\pi}N^{3/2}} \, 
\sqrt{\frac{(\Delta_1^2-1)(\Delta_2^2-1)(\Delta_3^2-1)}{
\Gamma\bigl(\tfrac{\Delta_1}{2}\bigr)
\Gamma\bigl(\tfrac{\Delta_1+3}{2}\bigr)
\Gamma\bigl(\tfrac{\Delta_2}{2}\bigr)
\Gamma\bigl(\tfrac{\Delta_2+3}{2}\bigr)
\Gamma\bigl(\tfrac{\Delta_3}{2}\bigr)
\Gamma\bigl(\tfrac{\Delta_3+3}{2}\bigr)}} \vspace*{5mm}  
\\
\vbox{\vskip1cm}
&\hspace*{5mm} \cdot
\frac{\Gamma\bigl(\tfrac{\Delta_1+\Delta_2-\Delta_3+2}{4}\bigr)
\Gamma\bigl(\tfrac{\Delta_2+\Delta_3-\Delta_1+2}{4}\bigr)
\Gamma\bigl(\tfrac{\Delta_3+\Delta_1-\Delta_2+2}{4}\bigr)}{
|\vec{x}_1-\vec{x}_2|^{\Delta_1+\Delta_2-\Delta_3}
|\vec{x}_2-\vec{x}_3|^{\Delta_2+\Delta_3-\Delta_1}
|\vec{x}_3-\vec{x}_1|^{\Delta_3+\Delta_1-\Delta_2}}
\langle C^{I_1} C^{I_2} C^{I_3}\rangle .
\end{split} \label{eq:three-point-function2}
\end{equation}

The expressions~\eqref{eq:three-point-function} and
\eqref{eq:three-point-function2} are certainly more formidable than
the expression found in~\cite{Lee:3point} for the analogous
correlation functions of the CPOs of the $\CN=4$, $D=4$ SCFT. 
One\footnote{We thank J.\ Distler for actually asking the question.}
can ask if these expressions simplify, given an explicit expression 
for the $\langle C^{I_1} C^{I_2} C^{I_3}\rangle$, which are related to
the Clebsch-Gordan coefficients for $SO(5)$. Since we know of no
closed-form expression for arbitrary values of $k$, let us consider
the very degenerate case that we have only the highest and lowest
weight states
\begin{equation}
\begin{split}
Y^{(k_1)} & = \frac{1}{2^{3k_1/2} \pi^{1/4}} (x_1+ix_2)^{k_1}, \\
Y^{(k_2)} & = \frac{1}{2^{3k_2/2} \pi^{1/4} }(x_1+ix_2)^{k_2}, \\
Y^{(k_3)} &= \frac{1}{2^{3k_3/2} \pi^{1/4}} (x_1-ix_2)^{k_3},
\end{split}
\end{equation}
where the normalization has been chosen to agree
with~\eqref{eq:Y-squared}. In this case, we find
\begin{equation}
\langle C^{(k_1)} C^{(k_2)} C^{(k_{3})}\rangle 
=\frac{\delta_{k_1+k_2,k_3}}{\pi^{1/4} 
\Gamma\bigl(k_1+k_2+\tfrac{5}{2} \bigr)}. \label{eq:c-c-c}
\end{equation}
For $k_3=k_1+k_2$, the 3-point function simplifies quite a bit, but
the use of~\eqref{eq:c-c-c} doesn't result in any further
simplification, 
\begin{equation}
\begin{split}
\langle  \CO^{I_1}(x_1) & \CO^{I_2}(x_2) \CO^{I_3} (x_3) \rangle \\
\vbox{\vskip1cm}
& = \frac{2^{(k_1+k_2-15)/2} \sqrt{\pi}}{N^{3/2}} 
\Gamma\bigl(k_1+\tfrac{1}{2} \bigr)
\Gamma\bigl(k_2+\tfrac{1}{2} \bigr)
\Gamma(k_1+k_2) 
\frac{\langle C^{(k_1)} C^{(k_2)} C^{(k_3)}\rangle }{
|\vec{x}_2-\vec{x}_3|^{4k_1}
|\vec{x}_3-\vec{x}_1|^{4k_2}} \\
\vbox{\vskip1cm}
& = \frac{2^{(k_1+k_2-15)/2} \pi^{1/4}}{N^{3/2}}  
\frac{\Gamma\bigl(k_1+\tfrac{1}{2} \bigr)
\Gamma\bigl(k_2+\tfrac{1}{2} \bigr)
\Gamma(k_1+k_2) }{
 \Gamma\bigl(k_1+k_2+\tfrac{5}{2} \bigr)} \,
\frac{1}{
|\vec{x}_2-\vec{x}_3|^{4k_1}
|\vec{x}_3-\vec{x}_1|^{4k_2}}.
\end{split}
\end{equation}
So, unfortunately, knowing the Clebsch-Gordan coefficients doesn't
appear to lead to more illuminating expressions for the correlation
functions. 

One further observation we can make about the 3-point functions is
that there don't appear to be any additional zeros beyond those
arising from the group-theoretic factor 
$\langle C^{I_1} C^{I_2} C^{I_3}\rangle$. This is, at least,
consistent with the fact that no additional zeros were found in an
analysis, based solely on the properties of the six-dimensional
superconformal algebra, of the 4-point
function~\cite{Distler:Six-SCFT}. 

From the $G^{\phi ss}_{I_1, I_2 I_3}$ vertex, we can also compute the
correlation function 
$\langle  \Phi^{I_1}\CO^{I_2} \CO^{I_3}\rangle$. 
We must first identify the normalization constant appearing in the
coupling of $\phi^I_0$ to $\Phi^I$ in the generating function. From
the 2-point function  
$\langle  \Phi^{I_1} \Phi^{I_2}\rangle$, we find
\begin{equation}
\CN^I_\Phi = -\frac{2^{(3k/2+1)}(2k+3)}{N^{3/2}}\,
\sqrt{\frac{(k+2)\Gamma\bigl(k+\tfrac{7}{2}\bigr)}{\Gamma(k+1)}}.
\end{equation}
We then compute, in the large $N$ limit,
\begin{equation}
\begin{split}
& \langle  \Phi^{I_1}(x_1) \CO^{I_2}(x_2) \CO^{I_3} (x_3) \rangle \\
\vbox{\vskip1cm}
 & =\frac{2^{11\Sigma/2-9}}{N^{3/2}} \, 
\frac{\Sigma(\Sigma+1)(\Sigma+3)\Gamma(\Sigma+6)}{
\Gamma\bigl(\tfrac{\Sigma+5}{2}\bigr)} 
\left[ \prod_{i=1}^3 
\frac{\Gamma(2\alpha_i+3)\Gamma\bigl(k_i+\tfrac{5}{2}\bigr)}{
\Gamma(\alpha_i+1)\Gamma(2k_i+4)}
\sqrt{\frac{\Gamma\bigl(k_i+\tfrac{3}{2}\bigr)}{
\Gamma(k_i)}} \right] \\
\vbox{\vskip1cm}
& \hspace*{5mm}\cdot
\alpha_1(\alpha_1-1)(2\alpha_1-3) \frac{2k_1+5}{k_1} 
\frac{(2k_2-3)(2k_3-3)}{(k_2-1)(k_3-1)} \,
\sqrt{(4k_2^2-1)(4k_3^2-1)} \\
\vbox{\vskip1cm}
& \hspace*{5mm}\cdot\frac{\langle T^{I_1} C^{I_2} C^{I_3}\rangle }{
|\vec{x}_1-\vec{x}_2|^{4\alpha_3+6}
|\vec{x}_2-\vec{x}_3|^{4\alpha_1-6}
|\vec{x}_3-\vec{x}_1|^{4\alpha_2+6}}.  \vspace*{5cm} 
\end{split}
\end{equation}
Once again, we can express this in terms of the conformal weights,
\begin{equation}
\begin{split}
& \langle  \Phi^{I_1}(x_1) \CO^{I_2}(x_2) \CO^{I_3} (x_3) \rangle \\
\vbox{\vskip1cm}
& = \frac{2^{(\Delta_1-\Delta_2-\Delta_3-23)/2}\pi}{N^{3/2}} \,
\frac{(\Delta_1+\Delta_2+\Delta_3-6)
(\Delta_1+\Delta_2-\Delta_3-12)
\Gamma\bigl(\tfrac{\Delta_1+\Delta_2+\Delta_3+6}{2}\bigr)
}{
(\Delta_2-3)(\Delta_3-3)(\Delta_2-2)(\Delta_3-2)
\Gamma\bigl(\tfrac{\Delta_1+\Delta_2+\Delta_3-4}{4}\bigr)}
\\
\vbox{\vskip1cm}
& \hspace*{5mm}\cdot \frac{\Gamma\bigl(\tfrac{\Delta_1}{2}-2\bigr)
\Gamma\bigl(\tfrac{\Delta_1+\Delta_2-\Delta_3}{2}\bigr)^2
\Gamma\bigl(\tfrac{\Delta_3+\Delta_1-\Delta_2}{2}\bigr)
}{
\Gamma(\Delta_1)
\Gamma\bigl(\tfrac{\Delta_2}{2}+2\bigr)
\Gamma\bigl(\tfrac{\Delta_3}{2}+2\bigr)
\Gamma\bigl(\tfrac{\Delta_1+\Delta_2-\Delta_3-2}{4}\bigr)
\Gamma\bigl(\tfrac{\Delta_1+\Delta_2-\Delta_3-10}{4}\bigr)
\Gamma\bigl(\tfrac{\Delta_3+\Delta_1-\Delta_2-2}{4}\bigr)
}
\\
\vbox{\vskip1cm}
& \hspace*{5mm}\cdot
\sqrt{
\frac{(\Delta_1-2)\Gamma\bigl(\tfrac{\Delta_1+1}{2}\bigr)
}{
\Gamma\bigl(\tfrac{\Delta_1}{2}-2\bigr)} \,
\frac{\Gamma\bigl(\tfrac{\Delta_2}{2}\bigr)
\Gamma\bigl(\tfrac{\Delta_2}{2}\bigr)
}{
(\Delta_2^2-1)(\Delta_3^2-1)\Gamma\bigl(\tfrac{\Delta_2+3}{2}\bigr)
\Gamma\bigl(\tfrac{\Delta_3+3}{2}\bigr)}}
\\
\vbox{\vskip1cm}
& \hspace*{5mm}\cdot\frac{\langle T^{I_1} C^{I_2} C^{I_3}\rangle }{
|\vec{x}_1-\vec{x}_2|^{\Delta_1+\Delta_2-\Delta_3}
|\vec{x}_2-\vec{x}_3|^{\Delta_2+\Delta_3-\Delta_1}
|\vec{x}_3-\vec{x}_1|^{\Delta_3+\Delta_1-\Delta_2}}.
\end{split} \label{eq:phi-cpo-cpo}
\end{equation}
As we saw for the case of
$\langle \CO^{I_1} \CO^{I_2} \CO^{I_3} \rangle$, we don't expect that
an explicit expression for  $\langle T^{I_1} C^{I_2} C^{I_3}\rangle$
will cause these expressions to simplify.

\section{The Operator Product Expansion of Wilson Surfaces}
\label{sec:surfaces}

In~\cite{Maldacena:Wilson}, it was shown that one could use the
AdS description of the large~$N$ limit of the $(0,2)$ superconformal
field theory in six dimensions to compute Wilson surface
observables~\cite{Ganor:largeN}. The chiral primaries for
these theories are
known~\cite{Aharony:AdS,Minwalla:PrimaryOps,Leigh:Large-N}, so one can
use the AdS formalism to write an operator product expansion for such
a surface. Analogously to the case of the Wilson
loop~\cite{Berenstein:OPE}, we expect that there exists an operator
product expansion for the Wilson surface when it is probed from
distances large compared to its characteristic size $a$,
\begin{equation}
W(\CS) = \langle W(\CS) \rangle \left[ 1 + \sum_{i,n} 
c^{(n)}_i a^{\Delta^{(n)}_i} \CO^{(n)}_i\right], \label{eq:wilssurfope}  
\end{equation}
where the $\CO^{(n)}_i$ are a set of operators with conformal weights
$\Delta^{(n)}_i$.  Here  $\CO^{(0)}_i$ denotes the
$i^{th}$ primary field, while the $\CO^{(n)}_i$ for $n>0$ are its conformal
descendants. For a spherical Wilson surface, the expectation
value of all operators other than the identity vanishes, so that the
coefficient of the identity is the expectation value of the loop. 

In~\cite{Berenstein:OPE} the spherical Wilson surface solution was
studied. The scalar charge of the surface was taken to be constant (a
point on $S^4$) and the Wilson surface was the 
boundary (a 2-sphere) of a  minimal area membrane worldvolume in
AdS$_7\times S^4$. A convenient parameterization of the solution was
given in terms of the Poincare  coordinates as
\begin{equation}
\begin{split} 
x_1&=\sqrt{a^2 - z^2} \cos\theta\\
x_2&=\sqrt{a^2 - z^2} \sin\theta\cos\psi\\
x_3&=\sqrt{a^2 - z^2} \sin\theta\sin\psi,
\end{split} \label{eq:surf-param}
\end{equation}
where $0\leq z \leq a$, $0\leq\theta\leq \pi$, and $0\leq \psi\leq 
2\pi$. 

The volume of the membrane was found to be divergent
\begin{equation}
\begin{split}
S &= T^{(2)} \int d {\cal V} =    
T^{(2)}  4 \pi \int_\epsilon^a \frac{ dz \, a \sqrt{a^2 - z^2 } }{ z^3}\\
& =  \pi T^{(2)}  \left[ 
+ \frac{2 a^2}{\epsilon^2 } - 2 \ln\frac{2 a }{\epsilon} -1 + 
{\cal O}(\epsilon)
 \right],
\label{eq:div-mem-act} 
\end{split}
\end{equation}
where the tension of the membrane is $T^{(2)}=2N/\pi$ in our units.
The quadratic divergence is proportional to the area of the surface
and was present in the case of a rectangular Wilson 
surface~\cite{Maldacena:Wilson}. The logarithmic divergence was found
to be proportional to the ``rigid string'' action
\cite{Polyakov:Fine-Struct}, where for a general 2-surface $\Sigma_2$, 
\begin{equation}
S_{rigid} = \int_{\Sigma_2}  d^2 \sigma  \sqrt{\gamma} ( \nabla^2 X^i)^2,
\label{eq:rigid} 
\end{equation}
where $\gamma$ is the induced metric on the Wilson surface and the
$X^i$ are the coordinates on $\BR^6$ describing the surface.  It was
conjectured in~\cite{Berenstein:OPE} that tensionless strings in the
$(0,2)$ six-dimensional field theory might be governed by some
supersymmetric form of the action~\eqref{eq:rigid}.  In fact, this
instance of a conformal anomaly is part of a much more general
structure, as discussed in~\cite{Graham:ConfAnom}.

One implication of the logarithmic divergence
in~\eqref{eq:div-mem-act} is that the expectation value
of the Wilson surface is not well defined, since we can add any constant
to the logarithmic subtraction. Furthermore, it seems to indicate that
the expectation value of a Wilson surface is scale dependent. 
Despite this, the connected correlation functions of
Wilson surfaces do not receive extra divergent
contributions. Therefore their correlators can be calculated in a
completely analogous fashion to the Wilson loops
in~\cite{Berenstein:OPE}, allowing the extraction of OPE coefficients
of~\eqref{eq:wilssurfope}. In this section, we investigate the OPE for
the spherical Wilson surface. We consider a 
Wilson surface whose characteristic size is much smaller than its
distance from any probe in the theory. Then we identify
the operators of low conformal dimension that are allowed to appear in
the OPE and we compute the necessary correlation functions to extract
the OPE coefficients. 

\subsection{The Operator Product Expansion of the Spherical Wilson
Surface} 

The possible operators which can appear in the 
Wilson surface must have the same symmetry properties as the Wilson
surface itself, so the operators $\CO^{(n)}_i$ should be bosonic and $S_{N+1}$
invariant.
The Wilson loop and surface solutions of~\cite{Maldacena:Wilson}
require a massive ``quark'' in the theory, and hence a non-zero Higgs
VEV. We consider the case where $\theta^I(s) = \theta^I$ is a
constant, which breaks the R-symmetry group from $SO(5)\rightarrow
SO(4)$. So we should look for operators in $SO(5)$ representations
which have $SO(4)$ singlets in their decomposition. We also consider
only operators whose dimensions are protected. The appearance in the
OPE of operators with large anomalous dimensions will
be suppressed.

The first level at which operators can appear is at dimension 
$\Delta= 4$, where we have the first CPO. This operator is in the 
$\mathbf{14}\rightarrow \mathbf{1}\oplus\mathbf{4}\oplus\mathbf{9}$
under $SO(4)$, so it is allowed. In general, the
CPOs, being in the symmetric traceless representations
of $SO(5)$, always contain one $SO(4)$ singlet, and hence all of them
may appear in the OPE.

At dimension $\Delta=5$, there are two operators to consider. There is a
$Spin(5,1)$ vector operator in the 
$\mathbf{10}\rightarrow\mathbf{4}\oplus\mathbf{6}$ which does not
contribute a singlet. There is also a Lorentz 3-form operator in the
$\mathbf{5}\rightarrow\mathbf{1}\oplus\mathbf{4}$ which is allowed to
contribute, depending on the orientation, $\hat{\sigma}^{\mu\nu\rho}$, of
the membrane. This is the first operator in a series of Lorentz
3-form, $SO(5)$ rank~$k$ symmetric traceless tensor operators of
dimension $\Delta=2k+3$, and all of these can appear. Below we denote
these operators as $\CO_{\mu\nu\rho}^I$. 

At dimension~$\Delta=6$, we have the CPO at $k=3$ which, as discussed
above, is permitted in the OPE. Additionally there is the first ($k=0$)
member of a series in the $\mathbf{20^\prime}$ of $Spin(5,1)$ and the
rank~$k$ symmetric traceless tensor of $SO(5)$, with dimensions
$\Delta=2k+6$.  

Higher dimension operators can be analyzed in the same fashion. 
We arrive at the following expression for the OPE
\begin{equation}
 \frac{ W(\CS) }{\langle W(\CS) \rangle }
 =  \Bid + c^{(0)}_2 \, a^4 \,   Y^I(\theta) \CO^I 
+ c^{(0)}_3 \, a^5 \, Y^I(\theta) 
\hat{\sigma}^{\mu\nu\rho} \CO^I_{\mu\nu\rho} +\cdots ,
\label{eq:low-ord-exp-sphere}
\end{equation} 
where the $Y^I(\theta)$ denote spherical harmonics on $S^4$ and the
coefficients $c^{(n)}_i$ are what we wish to compute.  

The coefficients of the operator
product expansion~\eqref{eq:low-ord-exp-sphere} can be determined in a
few ways. One can compute the correlator of the Wilson surface with
each operator that is expected to appear in the OPE. This correlator
gets contributions only from the given conformal primary and its
descendents. For a Wilson surface of size $a$, separated from an
operator by a distance~$L$, 
\begin{equation}
\frac{ \langle W(\CS) \CO^{(0)}_i \rangle }{ \langle W(\CS) \rangle }
= c^{(0)}_i \frac{a^{\Delta^{(0)}_i}}{L^{2\Delta^{(0)}_i}} 
+\sum_{m > 0} c^{(m)}_i a^{\Delta^{(m)}_i}
\langle \CO^{(m)}_i \CO^{(0)}_i \rangle, \label{eq:corr-singop}
\end{equation}
where we have assumed that the operators have been normalized.
Here we have isolated the contribution from the descendents and their
mixings with the primaries in the second term. One can also compute
the correlator of a pair of Wilson surfaces that are separated  
by a distance~$L$ which is large compared to their size. This correlator
can be calculated from the operator 
product expansion for the two Wilson surfaces
\begin{equation}
\begin{split}
\frac{ \langle W(\CS,L) W(\CS,0) \rangle }{ 
\langle W(\CS,L) \rangle \langle W(\CS,0) \rangle  }  = &
\sum_{i,j;m,n} c^{(m)}_i c^{(n)}_j \, 
a^{\Delta^{(m)}_i+\Delta^{(n)}_j}
\langle\CO^{(m)}_i(L)\CO^{(n)}_j(0)\rangle \\
= &  \sum_i \bigl(c^{(0)}_i\bigr)^2 \, 
\frac{a^{2\Delta^{(0)}_i}}{{L^{2\Delta^{(0)}_i}}} \\ 
& + \sum_{i,\{m,n\}\neq\{0,0\}} c^{(m)}_i c^{(n)}_i \, 
a^{\Delta^{(m)}_i+\Delta^{(n)}_i}
\langle\CO^{(m)}_i(L)\CO^{(n)}_i(0)\rangle.
\end{split} \label{eq:corr-twosurf}
\end{equation} 
In the last line, the first term is due solely to the primary fields,
while the second contains the contributions from descendents.

\subsection{Wilson Surface Correlators and OPE Coefficients from
Supergravity}  

We would now like to investigate the process of extracting the
large~$N$ values of the OPE coefficients by using the AdS/CFT
correspondence to compute correlation functions
like~\eqref{eq:corr-singop} or~\eqref{eq:corr-twosurf}. 
If we consider two closely separated Wilson surfaces, then,
analogously to the case of two Wilson loops~\cite{Gross:Aspects}, for
small enough separation, instead of having two separate membranes
forming the surfaces, a single membrane worldsheet with both surfaces
as its boundary will be the configuration of minimal surface area. As
the surfaces are separated beyond a critical distance, this worldsheet
will degenerate into a long, thin cylinder connecting two minimal
surfaces. In a worldsheet picture, this long, thin cylinder
would represent the bulk picture of propagation of light supergravity
states between the two membranes. 

This picture tells us that the correlation
functions~\eqref{eq:corr-singop} and~\eqref{eq:corr-twosurf} can be
calculated by treating the membrane as a source for the fields
propagating in anti-de Sitter spacetime and then computing the
effective action for the propagation of supergravity states from the
surface to the point where the other operator, or the other surface,
is located. This is, of course, the analog of the method used for the
Wilson loop in~\cite{Berenstein:OPE}. 

We would like to compute the coupling of the membrane to the bulk
supergravity fields. We will be content with considering only the
lightest scalar modes, which are the most relevant when the surfaces
are separated by a distance which is large compared to their size. The
lightest states include the modes $s^I$ from
section~\ref{sec:correlators}, corresponding to the appearance of the
CPOs in the OPE. We saw that these modes were
related to several components of the supergravity fluctuations
in~\eqref{eq:ads-fluctuations}. Couplings to the modes $b^I$ will be
through worldvolume fermions, and should be suppressed in the small
spacetime curvature limit that we are considering.
The couplings to  modes of the graviton can be obtained by linearizing
the membrane effective action  
\begin{equation}
\begin{split}
S_{\text{eff.}} &= T^{(2)} \int d^3\sigma 
\sqrt{ \det G_{mn} \partial_\alpha x^m \partial_\beta x^n } \\
&= T^{(2)} \int d^3\sigma 
\sqrt{ \det g_{mn} \partial_\alpha x^m \partial_\beta x^n } 
\left[ 1 
+ \frac{1}{2} ( g_{mn} \partial_\alpha x^m \partial_\beta x^n)^{-1}
h_{mn} \partial_\alpha x^m \partial_\beta x^n + \cdots \right].
\end{split}
\end{equation}
Since the membrane is living at a point on $S^4$, we are only
interested in the components $h^I_{\mu\nu}$, which,
from~\eqref{eq:s-dependence},  may be written in
terms of $s^I$ as
\begin{equation}
h^I_{\mu\nu} = -\frac{1}{14}g_{\mu\nu}  s^I 
+ \frac{3}{16k(2k+1)} \nabla_{(\mu}\nabla_{\nu)} s^I. 
\end{equation}

To aid in computing the derivatives, we can use an argument
from~\cite{Berenstein:OPE}, which was also used to
derive~\eqref{eq:bound-deriv}.  Namely, we are looking for the leading 
terms in the expansions in $a/L$ of~\eqref{eq:corr-singop}
and~\eqref{eq:corr-twosurf}. These leading contributions arise from
derivatives with respect to $z$ which act on the numerator of the
expansion into a Green's function, 
{\it e.\ g.},~\eqref{eq:bulk-value}. So we find 
\begin{equation}
h^I_{\mu\nu} \sim - \frac{1}{8} \left( g_{\mu\nu} + 3 \delta^z_\mu
\delta^z_\nu g_{zz} \right) s^I. 
\end{equation}
We find that the coupling to the membrane is given by
\begin{equation}
S_{s^I} = -\frac{3 T^{(2)}}{8} \int d\CV \, \frac{z^2}{a^2} s^I.
\end{equation}

The contributions to the correlator of two Wilson surfaces due to
exchange of $s^I$ modes are then expressed as
\begin{equation}
\frac{\langle W(\CS,L) W(\CS,0)\rangle^{(s)}}{
\langle W(\CS,L)\rangle \langle W(\CS,0)\rangle}
=\exp\left[ \sum_k \bigl[Y^I(\theta)\bigr]^2
\left(\frac{3T^{(2)}}{8}\right)^2 
\int d\CV\, d\CV^\prime \, \frac{(zz^\prime)^2}{a^4} G_k(W) 
\right].
\end{equation}
If the distance between the surfaces is much larger than their size,
then we may approximate the Green's function for propagation between
the two loops by the expression~\eqref{eq:UHS-Greens-approx}. We find
that 
\begin{equation}
\frac{\langle W(\CS,L) W(\CS,0)\rangle^{(s)}}{
\langle W(\CS,L)\rangle \langle W(\CS,0)\rangle}
\sim \sum_k \frac{2^{3k-1}\pi}{N} \, 
\frac{\Gamma(k)  }{\Gamma\bigl(k-\tfrac{1}{2}\bigr)} \, 
\bigl[Y^I(\theta)\bigr]^2
\left( \frac{a}{L}\right)^{4k}, \label{eq:surf-corr}
\end{equation}
where the $\sim$ refers to the fact that subleading terms at each
order of the $a/L$ expansion have been dropped but the numerical
coefficient corresponding to the coefficient of the contribution of
each primary field is precise.
We can similarly obtain the correlation function between the surface
and one of the chiral primary operators $\CO^I$,
\begin{equation}
\begin{split}
\frac{\langle W(\CS,L) \CO^I(0)\rangle^{(s)}}{
\langle W(\CS,L)\rangle \rangle}
\sim & \frac{1}{\CN^I} \frac{\delta}{\delta s_0^I(\vec{x})} Y^I(\theta) 
\int d\CV \, d^6x^\prime \left(
\frac{3T^{(2)}}{8}\,\frac{z^2}{a^2}\right)  
K_\Delta(\vec{x},z;\vec{x}^\prime) s_0^I(\vec{x}^\prime) \\
\sim & - 2^{(3k-1)/2} \,
\sqrt{\frac{\pi}{N}\, \frac{\Gamma(k)}{\Gamma\bigl(k-\tfrac{1}{2}\bigr)}} \,
 Y^I(\theta)
\frac{a^{2k}}{L^{4k}}, 
\end{split}
\end{equation}
where we have set the normalization of the operator by the 2-point
function~\eqref{largeN-2pt}. 
We note that the coefficient here is simply the square root of that
appearing in~\eqref{eq:surf-corr}.
From these correlation functions, we find the operator product
coefficients of the chiral primary operators 
\begin{equation}
c_{\Delta}^{\text{CPO}}
=-2^{(3\Delta-2)/4} \,
\sqrt{\frac{\pi}{N}\, 
\frac{\Gamma\bigl(\tfrac{\Delta}{2}\bigr)}{
\Gamma\bigl(\tfrac{\Delta-1}{2}\bigr)}},
\end{equation}
where the minus sign is a result of choosing the 3-point correlation
function of the $\CO^I$ to be positive.

\section{Conclusions}

In this paper we have succeeded in computing 3-point correlation
functions of chiral primary operators in the large $N$ limit of the
$(0,2)$ superconformal field theory. We also studied the operator
product expansion for a spherical Wilson surface, and found that we
could compute correlation functions of two surfaces as well as that
for a surface with a local operator. We used these correlation
functions to extract the explicit large $N$ values of operator product
coefficients for the chiral primary operators. 
Extracting more information about the $(0,2)$ theory from our results
is an exciting prospect to be left for future work. 

It would also be interesting to study the 3-point function from the
perspective of the DLCQ formulation of the $(0,2)$
theory using the prescription of~\cite{Aharony:02SCFT}. A discussion
of the comparison of DLCQ theories of M5-branes and the AdS/CFT
correspondence appears in~\cite{Awata:AdS7-Matrix}. In particular,
in the large $N$ limit of the DLCQ, there should be exact agreement
with our results.  

\section*{Acknowledgements}
We  would like to thank Ofer Aharony, David Berenstein, 
Philip Candelas,  Jacques Distler, Willy Fischler, 
Elie Gorbatov, Sangmin Lee, Juan Maldacena, and Moshe Rozali
for helpful discussions and correspondence. In addition, R.\ C.\ would 
like to thank D.\ Berenstein, W.\ Fischler, and J.\ Maldacena for
their collaboration in obtaining the results reported in
section~\ref{sec:surfaces}.

\appendix
%\stepcounter{section}

\section{Green's Functions and Correlation Functions}
\label{sec:greens}

In this appendix, we list some properties of Green's functions on AdS
which will be useful to us in the main text. We also discuss the
method we use to ensure that the limit $z\rightarrow 0$ is taken
properly when computing correlation functions. Scalar Green's
functions on anti-de Sitter space have been discussed 
in a large number of papers, 
including~\cite{Gubser:Correlators,Witten:Holography,Leigh:Large-N,%
Muck:Correlators,Freedman:Correlators,Chalmers:R-current}.

In the text, we consider a real scalar field $\phi$ on anti-de Sitter
spacetime of radius one. With source~$J$, the action and
equation of motion are
\begin{equation}
\begin{gathered}
S =- \int_{AdS} d^{d+1}x \, \sqrt{g}\, \left[ 
\frac{A}{2} \bigl( (\nabla\phi)^2 
+ m^2 \phi^2 \bigr) - \phi J \right], \\
A(-\nabla^2_x + m^2)\phi=J,
\end{gathered} \label{eq:action-eom}
\end{equation} 
where $A$ is some constant. Imposing the boundary
condition $\phi|_{\partial M} = 0$ yields a 
unique solution for $\phi$, as long as the operator 
$-\nabla^2_x + m^2$,
is positive definite. This is the case for all 
$m^2\geq -d^2/4$~\cite{Breitenlohner:PosEnergy}. 

Solutions for $\phi$ which minimize
the action~\eqref{eq:action-eom} are given by the integral equation
\begin{equation}
\phi(x) = \int_M d^{d+1}x^\prime \sqrt{g(x^\prime)} \, G(x,x^\prime)
J(x^\prime), \label{eq:phi-sol}
\end{equation}
where the kernel $G(x,x^\prime)$ is the covariant Green's function 
for the equation of motion~\eqref{eq:action-eom}, satisfying
\begin{equation}
A(-\nabla^2_x + m^2 )G(x,x^\prime) 
= \frac{1}{\sqrt{g(x^\prime)}}\delta^{(d+1)}(x-x^\prime).
\label{eq:greens-fn} 
\end{equation}
In the upper-half space representation of anti-de Sitter spacetime, with
metric 
\begin{equation}
ds^2 = \frac{1}{z^2} \left( dz^2 + \eta_{ij} dx^i dx^j \right), 
\label{eq:UHS-metric} 
\end{equation}
the scalar Green's function can only depend on the distance 
between the sources, 
\begin{equation}
W=\frac{z z^\prime}{(z - z^\prime)^2 +
\sum_{i=1}^d |x_i - x_i^\prime|^2}. 
\label{eq:wapprox}
\end{equation}
The solution for~\eqref{eq:greens-fn} which 
goes to zero at the boundary is given in terms of the hypergeometric 
function
\begin{equation}
G(W) = \frac{\alpha_0}{A}  W^{\Delta} \, 
{_2}F_1(\Delta,\Delta + \tfrac{1-d}{2},2\Delta-d+1;-4W). 
\label{eq:UHS-Greens} 
\end{equation}
where $\Delta = d/2 + \sqrt{ m^2 + d^2/4 }$ will be the conformal weight of 
the associated operator and $\alpha_0$ is 
\begin{equation}
\begin{split} 
\alpha_0 =& \frac{ \Gamma(\Delta)  }{
2 \pi^\frac{ d }{2} \Gamma(  \Delta -\tfrac{d}{2} +1 ) },
\\
= &  \frac{(\Delta-1)(\Delta-2)}{2 \pi^3 }
~~~~~~~~~~~~{\rm for}~~ d =6.
\end{split} \label{eq:alphanaught} 
\end{equation}

We will be particularly interested in fields in the bulk which are
produced by a source on the boundary. We take a source with support
very close to the boundary, 
$\text{supp}\bigl(J(x^\prime)\bigr)
= \{z^\prime | 0\leq z^\prime < \epsilon \}$. 
In this region $W\sim 0$, so we can approximate
\begin{equation}
G(W) \sim \frac{\alpha_0}{A}  W^{\Delta} . 
\label{eq:UHS-Greens-approx} 
\end{equation}
Then the field produced in the bulk is given by 
\begin{equation}
\begin{split}
\phi(x) & = \int_M d^{d+1}x^\prime \sqrt{g(x^\prime)} \, G(x,x^\prime)
J(x^\prime) \\
& \sim  \frac{\alpha_0}{A}   \int_{\partial M} d^{d}x^\prime
\left( \frac{z}{z^2+|\vec{x}-\vec{x}^\prime|^2} \right)^\Delta 
\int_0^\epsilon dz^\prime \, {z}^{\Delta-d-1}
J(x^\prime) \\
&= \int_{\partial M}  d^{d}x^\prime 
K_\Delta(\vec{x},z;\vec{x}^\prime) \phi_0(\vec{x}^\prime). 
\end{split} \label{eq:bulk-value}
\end{equation}
Here we have defined the boundary-to-bulk propagator
\begin{equation}
K_\Delta(\vec{x},z;\vec{x}^\prime) = 
\frac{ \Gamma(\Delta)  }{
A\pi^\frac{ d }{2} \Gamma(  \Delta -\tfrac{d}{2}) } 
\left( \frac{z}{z^2+|\vec{x}-\vec{x}^\prime|^2} \right)^\Delta 
\label{eq:boundary-to-bulk}
\end{equation}
and the boundary sources
\begin{equation}
\phi_0(\vec{x}^\prime) = \frac{1}{2\Delta-d} 
\int_0^\epsilon dz^\prime \, {z}^{\Delta-d-1} J(x^\prime).
\label{eq:bound-source}
\end{equation}

The normalization of the boundary-to-bulk
propagator~\eqref{eq:boundary-to-bulk} has been chosen specifically to
agree with that of~\cite{Freedman:Correlators}.  By evaluating the
action~\eqref{eq:action-eom} in terms of 
the boundary sources, 
\begin{equation}
\begin{split}
S^{(2)} &= \frac{1}{2} \int_M d^{d+1}x \, \int_{M^\prime}  d^{d+1}x^\prime \,
\sqrt{g(x)} \, J(x) \, G(x,x^\prime) \, \sqrt{g(x^\prime)} \,
J(x^\prime) \\
& = \frac{1}{2A} \,  \frac{2\Delta -d}{\Delta} \,
\frac{ \Gamma(\Delta+1)  }{
\pi^\frac{ d }{2} \Gamma(  \Delta -\tfrac{d}{2}) } 
\int d^{d}x \, d^{d}x^\prime \,
\frac{\phi_0(\vec{x})\phi_0(\vec{x}^\prime)}{
|\vec{x}_1-\vec{x}_1|^{2\Delta}}, 
\end{split} 
\end{equation}
we can compute the two-point function
\begin{equation}
\begin{split}
\langle \CO(x_1)\CO(x_2) \rangle
&= \frac{\delta^2S^{(2)}}{
\delta\phi_0(\vec{x}_1)\delta\phi_0(\vec{x}_2)} \\
&= \frac{1}{A} \,  \frac{2\Delta -d}{\Delta} \,
\frac{ \Gamma(\Delta+1)  }{
\pi^\frac{ d }{2} \Gamma(  \Delta -\tfrac{d}{2}) } 
\frac{1}{|\vec{x}_1-\vec{x}_2|^{2\Delta}},
\end{split} \label{eq:two-point-gen}
\end{equation}
in agreement with the ``corrected'' result found
in~\cite{Freedman:Correlators}. We note in
particular that the infamous factor of $(2\Delta-d)/\Delta$ which was
very carefully obtained in~\cite{Freedman:Correlators} appears here
very naturally. Of course, we are free to adjust the normalization
coefficient in~\eqref{eq:bound-source} so that the corresponding
operators are normalized. 

In the text, we are also interested in the three-point function
computed from a cubic term in the supergravity action of the form
\begin{equation}
L^{(3)} =  \sum_{I_1, I_2, I_3} G_{I_1 I_2 I_3} 
\phi^{I_1}\phi^{I_2}\phi^{I_3}.
\end{equation}
The computation of the three-point function from this vertex can be
found in~\cite{Freedman:Correlators}. Including the appropriate
combinatorial factor, we find
\begin{equation}
\begin{split}
\langle \CO^{I_1}(x_1) & \CO^{I_2}(x_2) \CO^{I_3} (x_3) \rangle
\vspace*{1mm} \\
= & \frac{3\Gamma\bigl(\tfrac{\Delta_1+\Delta_2-\Delta_3}{2}\bigr)
\Gamma\bigl(\tfrac{\Delta_2+\Delta_3-\Delta_1}{2}\bigr)
\Gamma\bigl(\tfrac{\Delta_3+\Delta_1-\Delta_2}{2}\bigr)  
\Gamma\bigl(\tfrac{\Delta_1+\Delta_2+\Delta_3}{2}-d\bigr)
}{
A_{I_1}A_{I_2}A_{I_3} \pi^d 
\Gamma(\Delta_1-\tfrac{d}{2}) \Gamma(\Delta_2-\tfrac{d}{2})
\Gamma(\Delta_3-\tfrac{d}{2})  } \vspace*{2mm} \\
& \cdot
\frac{
G_{I_1 I_2 I_3}
}{
|\vec{x}_1-\vec{x}_1|^{\Delta_1+\Delta_2-\Delta_3}
|\vec{x}_2-\vec{x}_3|^{\Delta_2+\Delta_3-\Delta_1}
|\vec{x}_3-\vec{x}_1|^{\Delta_3+\Delta_1-\Delta_2}}.
\end{split} \label{eq:three-point-form}
\end{equation}

\section{The Quadratic Action for the $s^I$ Modes}
\label{sec:quadaction}

In the text we need a Lagrangian for the modes $s^I$ up to cubic
order. In order to obtain the cubic $s$-$s$-$s$ vertex from the
quadratic equation of motion~\eqref{eq:lambda-s-s-s}, we needed to
know the normalization of the quadratic $s^I$ action. In
section~\ref{sec:correlators}, we were able to obtain this
normalization by comparison of the equations of
motion~\eqref{eq:phi-ss} and~\eqref{eq:sphi-s-eom}, used to extract
the  $s$-$s$-$\phi$ vertex.  
In~\cite{Lee:3point,Arutyunov:QuadAct} the computation of
the quadratic action for the $s^I$ modes on AdS$_5\times S^5$ was
performed (\!\cite{Arutyunov:QuadAct} actually considers the quadratic
action for {\it every} mode) using two different methods. 

We can first
attempt to use the procedure described in~\cite{Lee:3point}. Namely,
we expand the Lagrangian~\eqref{eq:11d-act} to quadratic order in all
fields which depend on $s^I$, as determined
by~\eqref{eq:s-dependence-coeffs}. We keep all terms in the expansion
of $\sqrt{-G}\,R$, including total
derivatives, and apply the equations of
motion whenever possible. To enforce the
constraint~\eqref{eq:rab-constraint}, we directly substitute the
solution for $h^\prime_{(\mu\nu)}$ given
by~\eqref{eq:solve-constraint}.  Via this method, we also succeeded in
recovering equation~(3.22) of~\cite{Lee:3point} when expanding around
the AdS$_5\times S^5$ background. In the present case, we find the
Lagrangian quadratic in the modes $s^I$ 
\begin{equation}
\begin{split}
L[s^I] = 
-\frac{2N^3z(k)}{\pi^5} &
\left( \frac{9(11k^2+21k-63)}{896k^2} (\nabla_\mu s^I)^2
+\frac{9}{224}(k-3)(11k+3)(s^I)^2 \right. \\
& + \frac{9(2k^2+6k-1)}{512k^2(2k+1)^2} 
(\nabla_{(\mu}\nabla_{\nu)}s^I)^2 \\
& \left. +\frac{9}{1024k^2(2k+1)^2} 
(\nabla_\lambda\nabla_{(\mu}\nabla_{\nu)}s^I)^2 \right),
\end{split} \label{eq:quad-action-raw}
\end{equation}
where $z(k)$ is the value of the integral over spherical harmonics
given in~\eqref{eq:double-harm-int-values}.

We wish to use the action~\eqref{eq:quad-action-raw} to compute the
2-point function of the CPOs. According to the
prescription of~\cite{Gubser:Correlators,Witten:Holography}, we should
compute the action as a function of boundary values. In order to do
this, we must carefully separate the boundary terms from the bulk in
the higher-derivative terms. 
For this, we compute 
\begin{equation}
\begin{split}
(\nabla_{(\mu}\nabla_{\nu)}s^I)^2 =& 
\nabla^\mu(\nabla^\nu s^I\nabla_\mu\nabla_\nu s^I) 
-2 ( 2k(k-3)-3 ) (\nabla_\mu s^I)^2 \\
& - \frac{16k^2(k-3)^2}{7}(s^I)^2, \\
(\nabla_\rho \nabla_{(\mu}\nabla_{\nu)}s^I)^2 
=& \nabla^\rho(\nabla^\mu\nabla^\nu s^I\nabla_\rho\nabla_\mu\nabla_\nu s^I) 
- 2 ( 2k(k-3)-7 ) \nabla^\mu(\nabla^\nu s^I\nabla_\mu\nabla_\nu s^I) \\
& + \frac{4}{7} \bigl( 4k(k-3)(6k(k-3)-35) +147 \bigr) (\nabla_\mu s^I)^2 \\
& -32 k^2(k-3)^2(s^I)^2,
\end{split} \label{eq:higher-deriv}
\end{equation}
on shell. We use these to rewrite~\eqref{eq:quad-action-raw} as
\begin{equation}
\begin{split}
L[s^I] = 
-\frac{2N^3z(k)}{\pi^5} & \left( \frac{9(k+1)(6k^2+2k+3)}{128k(2k+1)^2} 
\bigl[ (\nabla s^I)^2  +4k(k-3)(s^I)^2 \bigr] \right. \\
& + \frac{27}{256k^2(2k+1)} 
\nabla^\mu(\nabla^\nu s^I \nabla_\mu\nabla_\nu s^I)\\
& \left. + \frac{9}{1024 k^2(2k+1)^2}
\nabla^\rho(\nabla^\mu\nabla^\nu s^I 
\nabla_\rho\nabla_\mu\nabla_\nu s^I) \right) , 
\end{split} \label{eq:quad-action-bulk-vanishes}
\end{equation}
which demonstrates that the quadratic action vanishes on shell in the
bulk, as expected.

In order to finish the computation, we need to collect the surface
terms in~\eqref{eq:quad-action-bulk-vanishes} and compute them as a
function of boundary
values~\cite{Gubser:Correlators,Witten:Holography,Freedman:Correlators}
(see~\eqref{eq:bulk-value})  
\begin{equation}
s^I(z,\vec{x})= \frac{\Gamma(2k)}{A \pi^3 \Gamma(2k-3)}
\int d\vec{x}^\prime \, 
\left( \frac{z}{z^2+|\vec{x}-\vec{x}^\prime|} \right)^{2k} 
s^I_0(\vec{x}^\prime). \label{boundary-bulk}
\end{equation}
For this it is
convenient to use a prescription that was found useful
in~\cite{Berenstein:OPE}. We note that the leading singularities in
$1/|\vec{x}-\vec{x}^\prime|$ arise from terms which do not have
derivatives with respect to the boundary acting on $s^I$. Furthermore,
terms with $z$-derivatives acting on the denominator of the
boundary-to-bulk Green's function are also subleading. As only a
conformal structure is defined on the boundary of AdS, the induced
metric on the boundary is defined by rescaling by a power of
$z\rightarrow 0$. After this rescaling, these subleading terms will
vanish as the boundary is approached. Therefore, we only need to
compute the surface terms which involve 
\begin{equation}
\begin{split}
\partial_z s^I &\sim 2k s^I /z \\
\partial_z^2 s^I &\sim 2k(2k-1) s^I/z^2 . 
\end{split} \label{eq:bound-deriv}
\end{equation}
We use $\sim$ to indicate that terms which vanish in the limit that
$z\rightarrow 0$ have been dropped, but the numerical
coefficients are precise. 

We find that
\begin{equation}
\begin{split}
\nabla^\mu (\nabla^\nu s^I \nabla_\mu\nabla_\nu s^I) 
&\sim 4k^2 \nabla^\mu (s^I \nabla_\mu s^I) \\
\nabla^\rho (\nabla^\mu\nabla^\nu s^I \nabla_\rho \nabla_\mu\nabla_\nu s^I) 
&\sim 8k^2 ( 2k^2 + 3 ) \nabla^\mu (s^I \nabla_\mu s^I),
\end{split} \label{eq:higher-surface}
\end{equation}
so that
\begin{equation}
L[s^I] = 
- \frac{9(2k+3)z(k)}{128k} \, \frac{2N^3}{\pi^5} 
\nabla^\mu (s^I \nabla_\mu s^I).
\end{equation}
This is the same result that we would have obtained from a canonical
lagrangian (with no higher derivatives)
\begin{equation}
\begin{gathered}
L[s^I] = 
- \frac{A_I}{2} \bigl[ (\nabla s^I)^2  +4k(k-3)(s^I)^2 \bigr], \vspace*{2mm}
\\
A_I = \frac{9(2k+3)z(k)}{128k} \, \frac{2N^3}{\pi^5}.
\end{gathered} \label{eq:can-quad-action}
\end{equation}
Unfortunately,  this normalization is
inconsistent with the value~\eqref{eq:s-norm-true} obtained from  
the equations of motion of the theory at second order. In fact, it is
obvious that the $k$-dependence of this result is completely
incompatible with the totally symmetric form of the $s$-$s$-$s$ vertex
we obtain from~\eqref{eq:lambda-s-s-s}.

This procedure is also inconsistent in the case of
AdS$_5\times S^5$. While we reproduce the result~(3.22) (the 
higher derivative Lagrangian) of~\cite{Lee:3point}, when we evaluate
it as a function of boundary values
using the appropriate versions of
the formul\ae~\eqref{eq:higher-deriv} and~\eqref{eq:higher-surface},
we recover 
\begin{equation}
A^{(\text{AdS}_5)}_I = 32 k(k+2) z(k),  \label{eq:ads5-norm}
\end{equation}
 which differs by a
factor of $(k-1)/(k+1)$ from the result quoted as eq.~(3.23)
of~\cite{Lee:3point}. Nevertheless, the $k$-dependence of the
result reported there is certainly correct,
as it led to the cyclic symmetry of the 3-point vertex there.

We conclude that this method of evaluating the quadratic action is
flawed. 

The authors of~\cite{Arutyunov:QuadAct} note that it is conventional
to supplement the Einstein-Hilbert action with boundary terms
involving the second fundamental form of the metric $G$ 
as well as the metric induced on the
boundary~\cite{York:ConformalThree,Gibbons:Action-Integrals,%
Witten:Holography,Liu:Conformal,Arutyunov:Boundary}. 
In our case, the contribution of the standard extrinsic curvature term
can be easily computed and its  addition to 
the action above does not result in a  correct
normalization. 

The motivation of~\cite{Gibbons:Action-Integrals} (see an extended
discussion in~\cite{Hawking:PathIntegral}) to add the extrinsic
curvature term was to remove terms in the Einstein-Hilbert action that
were linear in second derivatives. As mentioned
in~\cite{Liu:Conformal,Arutyunov:Boundary}, this occurs for the 
physical graviton because certain terms in the Einstein-Hilbert 
action vanish if an appropriate gauge choice is made. Here we are 
dealing with the action for scalars that arises from the supergravity 
action and the terms in question cannot be made to vanish. Therefore 
adding the extrinsic curvature term does not cancel all of the 
boundary terms in the action for the scalar modes. 

Another motivation for adding boundary terms will arise
from supersymmetry. The supergravity action is supersymmetric up to
total derivative terms, so we should expect that additional boundary
terms must be added via the Noether procedure to restore supersymmetry
on a space with boundary\footnote{We thank Jacques Distler for
reminding us of this.}.  The process of finding all such terms seems
to be a much more difficult procedure than the method we use in the
text, so we will not attempt to do that here.

Despite these remarks, the authors of~\cite{Arutyunov:QuadAct} seem to
obtain the correct normalization for the $s^I$ action in the case of
AdS$_5\times S^5$. Applying what we understand their methods to be did
not result in the correct normalization for AdS$_7\times S^4$. 

\section{Quadratic Corrections to the Equations of Motion}
\label{sec:quadcorr}
 
We must compute quadratic corrections $Q^I_i$ 
to~\eqref{eq:rab-field1}, \eqref{eq:rab-constraint},
and~\eqref{eq:maxwell-field},  
\begin{equation}
-\frac{1}{8} \left[  
\left( \nabla^\mu\nabla_\mu 
+\frac{1}{10} \nabla^\gamma \nabla_\gamma
-72 \right) h_2^I 
+ \nabla^\gamma \nabla_\gamma h^{\prime I} 
- 192\sqrt{2} \nabla^\gamma \nabla_\gamma b^I \right] Y^I
+ Q^I_1  Y^I =0, 
\end{equation}
\begin{equation}
\left[ -\frac{1}{2} \left(  h^{\prime I} - \frac{9}{10} h_2^I \right)
+ Q^I_2 \right] 
\nabla_{(\alpha}\nabla_{\beta)} Y^I=0.
\end{equation}
\begin{equation}
-6 b^I \nabla_\alpha \nabla^\gamma\nabla_\gamma Y^I
+ \left[ -6 \left(\nabla^\mu\nabla_\mu   b^I 
- \frac{1}{2\sqrt{2}} h^{\prime I}
+\frac{3\sqrt{2}}{5} h_2^I \right) 
+ Q_3^I \right] \nabla_\alpha Y^I=0. 
\end{equation}
We then find that the equation of motion for the $s^I$ takes the form 
\begin{equation}
\left( \nabla^\mu\nabla_\mu - 4k(k-3)\right) s^I = 
\frac{8k}{2k+3} \left( Q_1^I +(k+3)(k+4) Q_2^I 
+\frac{2\sqrt{2} (k+3)}{3} Q_3^I \right). \label{eq:eom-s}
\end{equation}
We also compute quadratic equations to the field
equation~\eqref{eq:rab-field2} for the $\phi^I$ modes
\begin{equation}
\left[ -\frac{1}{2} \left( \nabla^\mu\nabla_\mu  
+ \nabla^\gamma \nabla_\gamma - 8 \right) \phi^I + Q^I_\phi \right] 
Y^I_{(\alpha\beta)} =0,  
\end{equation}

The terms $Q^I_1$ and $Q^I_2$ are determined by computing the
quadratic corrections 
to the $\alpha\beta$ components of~\eqref{eq:11d-eom}. For the $ss$
terms, using~\eqref{eq:s-dependence}, we find that  
\begin{equation}
\begin{gathered}
\delta^2 R_{\alpha\beta} |_{ss} = Z_{\alpha\beta} + g_{\alpha\beta} Y, \\
\delta^2\left[ \frac{1}{6} 
\left( F_{\alpha  m_1\cdots m_3}{F_\beta}^{m_1\cdots m_3} 
- \frac{1}{12} g_{\alpha\beta} F^2 \right)\right]_{ss} =
\zeta_{\alpha\beta} + g_{\alpha\beta} \psi,
\end{gathered}
\end{equation}
where
\begin{equation}
\begin{split}
Z_{\alpha\beta}  = & \frac{1}{4} \Bigl[  \left( 7 U_2 U_3 + 2 V_2 V_3 \right) 
\left( \nabla_\alpha s_2 \nabla_\beta s_3 +
2 s_2 \nabla_\alpha \nabla_\beta s_3 \right)
+ 2 V_2 \left( 7U_3+2V_3\right) 
\nabla_\alpha s_2 \nabla_\beta s_3 \vspace*{1mm}\\
&  \hspace*{1cm} +W_2 W_3 \left( 
\nabla^{(\mu}\nabla^{\nu)} \nabla_\alpha s_2 
\nabla_{(\mu}\nabla_{\nu)} \nabla_\beta s_3 +
2 \nabla^{(\mu}\nabla^{\nu)} s_2 
\nabla_{(\mu}\nabla_{\nu)} \nabla_\alpha \nabla_\beta s_3 \right) 
\Bigr], \vspace*{1mm}\\
Y = &\frac{1}{2} \Bigl[ U_2 V_3 \nabla^\mu (s_2\nabla_\mu s_3) 
+W_2 V_3 \nabla^\mu (\nabla_{(\mu}\nabla_{\nu)}s_2\nabla^\nu s_3)
+V_2V_3  \nabla^\gamma (s_2 \nabla_\gamma s_3) \\
& \hspace*{1cm}-\frac{V_2}{2} (7U_3+2V_3) \left( 
\nabla^\mu s_2 \nabla_\mu s_3 +\nabla^\gamma s_2 \nabla_\gamma s_3
\right) \Bigr],
\end{split}
\end{equation}
and
\begin{equation}
\begin{split}
\zeta_{\alpha\beta} = & 
-36 X_2 X_3 \nabla^\mu \nabla_\alpha s_2
\nabla_\mu\nabla_\beta s_3, \\
\psi = & 24 X_2 X_3 \left( \nabla^\gamma \nabla_\gamma s_2 
\nabla^\delta \nabla_\delta s_3 
+ \nabla^\mu \nabla^\gamma s_2
\nabla_\mu\nabla_\gamma s_3 \right) \\
& +72\sqrt{2} V_2 X_3 s_2 \nabla^\gamma \nabla_\gamma s_3  
+72 V_2 V_3 s_2 s_3.
\end{split}
\end{equation}
Here $U_2=U(k_2)$, etc., are the coefficients defined
in~\eqref{eq:s-dependence-coeffs}, while $s_2=s^{I_2}$, etc.

The term $Q^I_3$ is obtained from the quadratic terms in the 
Maxwell equation. For the $ss$ terms we compute
\begin{equation}
\begin{split}
\delta^2 (\nabla^m F_{m \alpha_1 \alpha_2 \alpha_3} )|_{ss}
=  & {\epsilon^{\delta}}_{\alpha_1 \alpha_2 \alpha_3} 
\Bigl[  W_2 \left( 6X_3 - \frac{3\sqrt{2}}{2} W_3  \right) 
\nabla^{(\mu}\nabla^{\nu)}  s_2 
\nabla_{(\mu}\nabla_{\nu)} \nabla_\delta s_3 \\
& \hspace*{1cm}+  3 \left( -5U_2 + 2V_3 
+ 12 W_2 \left(\frac{m_2^2}{7}-1\right) \right) X_3
\nabla^{\mu}  s_2 \nabla_{\mu} \nabla_\delta s_3 \\
& \hspace*{1cm}+ \left( 
6(U_2 m_3^2-V_2f_3)X_3
-3\sqrt{2} (7U_2 U_3 -10V_2V_3) \right. \\
& \hspace*{1.5cm} \left.
-18 X_2f_2 V_3
\right) s_2 \nabla_\delta s_3 \Bigr].
\end{split} \label{eq:ss-max}
\end{equation}

Computation of the projections onto spherical harmonics is facilitated
by the integral identities in
appendix~\ref{sec:sphere-harm}. Projection of 
$\tfrac{1}{4} (Z-\zeta)^\gamma_\gamma + Y-\psi$ onto
$Y^I$ yields the correction to~\eqref{eq:rab-field1},
\begin{equation}
\begin{split}
Q_1^{I_1} =  \frac{1}{4z_1} &
\Bigl[ \tfrac{1}{4}
\left( W_2 W_3 b_{123} 
+ \tfrac{8(W_2 V_3+ V_2 W_3) -W_2 W_3( f_2+f_3)}{2}
a_{123} \right)
\nabla^{(\mu}\nabla^{\nu)} s_2 \nabla_{(\mu}\nabla_{\nu)} s_3  \\
& +  \left( -60X_2 X_3\, b_{123}  
+ \tfrac{7( U_2 V_3 + V_2 U_3 ) + 6 \bigl( W_2 (m_2^2-7) V_3 
+ V_2 W_3 (m_3^2-7)\bigr)}{7} a_{123}  
\right) \nabla^\mu s_2 \nabla_\mu s_3  \\
&   +\Bigl(
\left( 
- 96(3 V_2 V_3 + X_2 X_3 f_2 f_3)
+  (U_2 V_3 m_3^2 + V_2 m_2^2 U_3 ) \phantom{\tfrac{7U}{4}}
\right. \\
& \hspace*{1cm} \left. + 144  \sqrt{2} (V_2 X_3 f_3 + X_2 f_2 V_3)   
-\tfrac{(7U_2 U_3 +6V_2 V_3) (f_2+f_3)}{4}  
\right)  a_{123} \\
& \hspace*{1cm} +  (7U_2 U_3+10V_2 V_3) b_{123}  
 \Bigr) s_2 s_3  
\Bigr]\langle C^{I_1}C^{I_2}C^{I_3}\rangle.
\end{split}
\end{equation}

Projecting $(Z-\zeta)_{(\alpha\beta)}$ onto 
$\nabla^{(\alpha}\nabla^{\beta)}Y^I$ yields the correction
to~\eqref{eq:rab-constraint}, 
\begin{equation}
\begin{split}
Q_2^{I_1} = \frac{1}{ 4q(k_1)f(k_1)z(k_1)} 
& \left[ W_2 W_3 ( c_{123} + d_{213}+d_{321}) 
\nabla^{(\mu} \nabla^{\nu)} s_2 
\nabla_{(\mu} \nabla_{\nu)} s_3 \right. \\
& + 144 X_2 X_3 c_{123} \nabla^\mu s_2 \nabla_\mu s_3 \\
& \left. +  (7U_2 U_3 + 2V_2 V_3)( c_{123} + d_{213}+d_{321})  s_2 s_3   
\right]\langle C^{I_1}C^{I_2}C^{I_3}\rangle.
\end{split}
\end{equation}

Projecting the terms in~\eqref{eq:ss-max} onto
${\epsilon^\delta}_{\alpha_1\alpha_2\alpha_3} \nabla_\delta Y^I$, we
find 
\begin{equation}
\begin{split}
Q_3^{I_1} = & \frac{b_{213}}{f_1 z_1} 
\Bigl[ \left( 3 ( W_2 X_3 + X_2 W_3 ) 
- \tfrac{3\sqrt{2}}{2} W_2 W_3  \right) 
\nabla^{(\mu}\nabla^{\nu)}  s_2 
\nabla_{(\mu}\nabla_{\nu)} s_3 \\
& \hspace*{1cm}+ \left( 
\tfrac{-15(U_2 X_3 + X_2 U_3) 
+6(V_2 X_3+ X_2 V_3 )}{2}
+\tfrac{36 (W_2 (m_2^2-7)X_3 + X_2 W_3 (m_2^3-7) )}{7} \right)
\nabla^{\mu}  s_2 \nabla_{\mu}  s_3 \\
& \hspace*{1cm}
+ \left(
-\tfrac{3\sqrt{2}}{2} (7U_2 U_3 -10V_2V_3) 
+3(U_2X_3+ X_2 U_3) 
\right. \\
& \hspace*{1.5cm} \left.  \phantom{\tfrac{3\sqrt{2}}{2}}
-12 (X_2 f_2 V_3 + V_2 X_3 f_3) \right) s_2  s_3 \Bigr] 
\langle C^{I_1}C^{I_2}C^{I_3}\rangle. 
\end{split}
\end{equation}

Projecting $(Z-\zeta)_{(\alpha\beta)}$ onto the symmetric tensor
spherical harmonics $Y^I_{(\alpha\beta)}$ yields the correction
to~\eqref{eq:rab-field2},
\begin{equation}
\begin{split}
Q_\phi^{I_1} = & \frac{h_{123}}{4z_1} 
\Bigl[  - W_2 W_3  
\nabla^{(\mu}\nabla^{\nu)}  s_2 
\nabla_{(\mu}\nabla_{\nu)} s_3 
+ 144 X_2 X_3 \nabla^{\mu}  s_2 \nabla_{\mu}  s_3 \\
& \hspace*{1cm} 
-( 7U_2 U_3 +2V_2 V_3) s_2  s_3 \Bigr]\langle T^{I_1}C^{I_2}C^{I_3}\rangle. 
\end{split} \label{eq:Qphi}
\end{equation}

For the $s\phi$ terms in the $\alpha\beta$ components
of~\eqref{eq:11d-eom}, we obtain
\begin{equation}
\begin{gathered}
\delta^2 R_{\alpha\beta} |_{s\phi} = Z^\phi_{\alpha\beta} 
+ g_{\alpha\beta} Y^\phi, \\
\delta^2\left[ \frac{1}{6} 
\left( F_{\alpha  m_1\cdots m_3}{F_\beta}^{m_1\cdots m_3} 
- \frac{1}{12} g_{\alpha\beta} F^2 \right)\right]_{s\phi} =
\zeta^\phi_{\alpha\beta} + g_{\alpha\beta} \psi^\phi,
\end{gathered}
\end{equation}
where
\begin{equation}
\begin{split}
Z^\phi_{\alpha\beta}  = &
-\frac{5U}{4} \nabla^\mu h_{(\alpha\beta)} \nabla_\mu s 
+\frac{U}{2}\nabla^\mu  \nabla_\mu h_{(\alpha\beta)} s  
+\frac{V}{2}\left(s \nabla^\gamma  \nabla_\gamma h_{(\alpha\beta)} 
+ \nabla^\gamma s \nabla_\gamma h_{(\alpha\beta)} 
\right) \\ &
 +\frac{W}{2} \nabla_\mu 
(\nabla_\nu h_{(\alpha\beta)} \nabla^{(\mu} \nabla^{\nu)} s) 
+\frac{7U}{4}(\nabla_\alpha h_{(\beta\gamma)}
+\nabla_\beta h_{(\alpha\gamma)} - \nabla_\gamma h_{(\alpha\beta)} )
\nabla^\gamma s \\
& -\frac{V}{2} (h_{(\alpha\gamma)} \nabla^\gamma\nabla_\beta s
+ h_{(\beta\gamma)}\nabla^\gamma\nabla_\alpha s )
-16V h_{(\alpha\beta)} s, \vspace*{1mm} \\
Y^\phi = & \frac{V}{2} h_{(\gamma\delta)} \nabla^\gamma\nabla^\delta s,
\end{split}
\end{equation}
and
\begin{equation}
\begin{split}
\zeta^\phi_{\alpha\beta} = & 
-24(\sqrt{2} X \nabla^\gamma \nabla_\gamma s + 2  V s)
h_{(\alpha\beta)}, \vspace*{1mm} \\
\psi^\phi = & 0.
\end{split}
\end{equation}
The $s\phi$ corrections to the Maxwell equations are
\begin{equation}
\begin{split}
\delta^2 (\nabla^m F_{m \alpha_1 \alpha_2 \alpha_3} )|_{s\phi}
=  & {\epsilon^{\delta}}_{\alpha_1 \alpha_2 \alpha_3} 
\Bigl[ \bigl( 12\sqrt{2} V - 6Xf  \bigr) 
h_{(\gamma\delta)} \nabla^\gamma s \Bigr] \\
& + {\epsilon^{\gamma\delta}}_{\alpha_2 \alpha_3} \Bigl[
18X \nabla^\mu h_{(\alpha_1 \delta)} \nabla_\mu \nabla_\gamma s
+\bigl( 18\sqrt{2}V-18Xf  \bigr)s \nabla_\gamma h_{(\alpha_1 \delta)}
\\
& \hspace*{2cm}
+9\sqrt{2}V h_{(\alpha_1\delta)} \nabla_\gamma s \Bigr].
\end{split} \label{eq:sphi-max}
\end{equation}

Projecting these expressions onto the appropriate spherical harmonics,
we find 
\begin{equation}
Q^{\phi {I_1}}_1 = -\frac{V_3 h_{213}}{4z_1}  \phi_2 s_3
\langle C^{I_1}T^{I_2}C^{I_3}\rangle, 
\end{equation}
\begin{equation}
\begin{split}
Q^{\phi {I_1}}_2 = & \frac{h_{213}}{q_1f_1z_1} \Bigl[ 
- \frac{W_3}{2}  \nabla^{(\mu} \nabla^{\nu)} \phi_2 
\nabla_{(\mu} \nabla_{\nu)} s_3 
+ \left( \frac{5 U_3}{4} 
- 3W_3 \left(\frac{m_3^2}{7}-1\right) 
\right) \nabla^\mu \phi_2 \nabla_\mu s_3 \\
&\hspace*{1cm} 
+  \left( 24\sqrt{2} X_3 f_3 
+\left(7+ \frac{7}{8}f_1+\frac{17}{8}f_2-\frac{7}{8}f_3\right)U_3
\right. \\
&\hspace*{1.5cm} \left. 
+\left(-34 +\frac{3}{4} f_2 -\frac{3}{4} f_3 \right) V_3 \right)
\phi_2 s_3 \Bigr]\langle C^{I_1}T^{I_2}C^{I_3}\rangle, 
\end{split}
\end{equation}
\begin{equation}
\begin{split}
Q^{\phi {I_1}}_3 = & \frac{3h_{213}}{f_1z_1} \Bigl[
-2 X_3 \nabla^\mu \phi_2 \nabla_\mu s_3
+ (3\sqrt{2} V_3 - 2 X_3 f_3) \phi_2 s_3 
\Bigr]\langle C^{I_1}T^{I_2}C^{I_3}\rangle.
\end{split}
\end{equation}

\section{Spherical Harmonics on $S^4$}
\label{sec:sphere-harm}

Here we collect some results about the spherical harmonics on $S^4$,
following the appendix of~\cite{Lee:3point}.  We take the radius of
the sphere to be $1/2$, as in the text. We embed the 
$S^4\subset \BR^5$ and represent the spherical harmonics by
\begin{equation}
Y^I = C^I_{i_1\cdots i_k} x^{i_1}\cdots x^{i_k}, 
\end{equation}
where we normalize 
$C^{I_1}_{i_1\cdots i_k} C^{I_2\, i_1\cdots i_k}
=\delta^{I_1 I_2}$. 
For each $k$, there are 
\begin{equation}
n(k)=(2k+3)(k+2)(k+1)/6
\end{equation}
harmonics and the eigenvalues of the spherical harmonics on a sphere
of radius $1/2$ are given by
\begin{equation}
\nabla^\alpha \nabla_\alpha Y^I = - 4k(k+3) Y^I. \label{eq:sh-eigen}
\end{equation}

Integrals over products of spherical harmonics are facilitated by the
use of the formula 
\begin{equation}
\int_{S^4} x^{i_1} \cdots x^{i_{2m}} = 
\frac{3\omega_4}{
2^{3m+2} \Gamma(m+\tfrac{5}{2})}
\bigl(  
\delta^{i_1i_2}\cdots\delta^{i_{2m-1} i_{2m}} + \text{perms.}\bigr),
\end{equation}
where $\omega_4=8\pi^2/3$ and the permutations are computed in the
following manner: fix the first index, then perform
$(2m-1)$ cyclic permutations, then, for each of the terms thus
generated, fix the first three indices and perform $(2m-3)$ cyclic
permutations, etc. We find the integral formul\ae
\begin{equation}
\begin{gathered}
\int_{S^4} Y^{I_1} Y^{I_2} = z(k) \delta^{I_1 I_2},
\label{eq:Y-squared} \\
\int_{S^4} \nabla^\alpha Y^{I_1} \nabla_\alpha Y^{I_2} 
= f(k) z(k) \delta^{I_1 I_2},  \\
\int_{S^4} \nabla^{(\alpha}\nabla^{\beta)} Y^{I_1} 
\nabla_{(\alpha}\nabla_{\beta)} Y^{I_2} = 
q(k) f(k) z(k) \delta^{I_1 I_2} ,
\end{gathered}
\end{equation}
where
\begin{equation}
\begin{gathered}
z(k) = \frac{3\omega_4\Gamma(k+1)}{2^{3k+2} \Gamma(k+\tfrac{5}{2})} , \\
f(k) = 4k(k+3), \\
q(k) = 3 (k-1)(k+4).
\end{gathered} \label{eq:double-harm-int-values}
\end{equation}
For any given $k$, the sum over the $n(k)$ degenerate eigenfunctions,
labeled by $A$, is a constant
\begin{equation}
\sum_A \left[ Y^I_A(\theta) \right]^2 =
\frac{(2k+3)\Gamma(k+3)}{2^{3(k+1)}
\Gamma(k+\tfrac{5}{2})}. \label{eq:harm-sum} 
\end{equation}

Further, we find
\begin{equation}
\begin{gathered}
\int_{S^4} Y^{I_1} Y^{I_2} Y^{I_3} = 
a(k_1,k_2,k_3) \langle C^{I_1}C^{I_2}C^{I_3}\rangle, \\
a(k_1,k_2,k_3) = \frac{3\omega_4 }{
2^{3\Sigma/2+2} \Gamma(\tfrac{\Sigma+5}{2})} 
\frac{\Gamma(k_1+1)\Gamma(k_2+1)\Gamma(k_3+1)}{
\Gamma(\alpha_1+1)\Gamma(\alpha_2+1)\Gamma(\alpha_3+1)},
\end{gathered} \label{eq:triple-int}
\end{equation}
\begin{equation}
\begin{gathered}
\int_{S^4} Y^{I_1} \nabla^\alpha Y^{I_2} \nabla_\alpha Y^{I_3} =  
b(k_1,k_2,k_3) \langle C^{I_1}C^{I_2}C^{I_3}\rangle, \\ 
b(k_1,k_2,k_3)  
= \frac{1}{2} \left( f(k_2) + f(k_3) - f(k_1) \right) 
a(k_1,k_2,k_3),
\end{gathered} 
\end{equation}
\begin{equation}
\begin{gathered}
\int_{S^4} \nabla^{(\alpha}\nabla^{\beta)} Y^{I_1} 
\nabla_\alpha Y^{I_2} \nabla_\beta Y^{I_3} = c(k_1,k_2,k_3)
\langle C^{I_1}C^{I_2}C^{I_3}\rangle, \\
c(k_1,k_2,k_3) = \frac{1}{4} 
\left( f(k_1)^2 - f(k_2)^2 - f(k_3)^2 + 2 f(k_2) f(k_3) \right)
a(k_1,k_2,k_3)
\end{gathered} 
\end{equation}
\begin{equation}
\begin{gathered}
\int_{S^4} Y^{I_1} \nabla^{(\alpha}\nabla^{\beta)} Y^{I_2} 
\nabla_{(\alpha}\nabla_{\beta)} Y^{I_3} =
d(k_1,k_2,k_3)
\langle C^{I_1}C^{I_2}C^{I_3}\rangle , \\
d(k_1,k_2,k_3) = \frac{1}{2} 
\left( f(k_2) + f(k_3) - f(k_1)-24 \right)  
b(k_1,k_2,k_3) \\
- \frac{f(k_2) f(k_3)}{4} a(k_1,k_2,k_3),
\end{gathered} 
\end{equation}
where $\langle C^{I_1}C^{I_2}C^{I_3}\rangle$, $\alpha_i$, and $\Sigma$
are 
defined as in~\eqref{eq:3C-contract}.

Rank two, symmetric traceless tensor spherical harmonics are defined
via
\begin{equation}
Y^{I}_{(\alpha\beta)}= 
e_{\alpha a} e_{\beta b} {T^{I \, ab}}_{i_1\cdots i_k}
x^{i_1} \cdots x^{i_k}, 
\end{equation}
and satisfy
\begin{equation}
\nabla^\gamma \nabla_\gamma Y^{I}_{(\alpha\beta)}
= -4(k(k+3)-2) Y^{I}_{(\alpha\beta)}. \label{eq:tensor-eigen}
\end{equation}
We have the integral formul\ae
\begin{equation}
\begin{gathered}
\int_{S^4} Y^{I_1}_{(\alpha\beta)} Y^{I_2}_{(\gamma\delta)}
g^{\alpha\gamma} g^{\beta\delta} 
= z(k) \langle T^{I_1} T^{I_2} \rangle, \\
\langle T^{I_1} T^{I_2} \rangle = 
{T^{I_1 \, ab}}_{i_1\cdots i_k} {T^{I_2}_{ab}}^{i_1\cdots i_k},
\end{gathered} 
\end{equation}
\begin{equation}
\begin{gathered}
\int_{S^4} Y^{I_1}_{(\alpha\beta)} Y^{I_2}_{(\gamma\delta)}
g^{\alpha\gamma} g^{\beta\delta} Y^{I_3}
= a(k_1,k_2,k_3) \langle T^{I_1} T^{I_2} C^{I_3} \rangle , \\
\langle T^{I_1} T^{I_2} C^{I_3} \rangle = 
{T^{I_1 \, ab}}_{i_1\cdots i_{\alpha_2+\alpha_3-1}} 
{{{T^{I_2}}_{ab}}^{i_1\cdots i_{\alpha_3}}}_{j_1\cdots j_{\alpha_1-1}} 
C^{I_3 i_{\alpha_3+1} \cdots i_{\alpha_3+\alpha_2-1}
j_1\cdots j_{\alpha_1-1}},
\end{gathered} 
\end{equation}
\begin{equation}
\begin{gathered}
\int_{S^4} Y^{I_1}_{(\alpha\beta)} \nabla^\alpha Y^{I_2} 
\nabla^\beta Y^{I_3} 
= h(k_1,k_2,k_3) \langle T^{I_1} C^{I_2} C^{I_3} \rangle, \\
\langle T^{I_1} C^{I_2} C^{I_3} \rangle = 
T^{I_1\, abi_1\cdots i_{\alpha_2+\alpha_3}} 
C^{I_2}_{ai_1\cdots i_{\alpha_3}j_1\cdots j_{\alpha_1-1}}
{C^{I_3}_{bi_{\alpha_3+1}
\cdots i_{\alpha_3+\alpha_2}}}^{j_1\cdots j_{\alpha_1-1}} , \\
h(k_1,k_2,k_3) = (\Sigma+3)\alpha_1 a(k_1,k_2,k_3),
\end{gathered} 
\end{equation}
\begin{equation}
\begin{gathered}
\int_{S^4} Y^{I_1}_{(\alpha\beta)} 
\nabla^\gamma \nabla^{\alpha} Y^{I_2} \nabla_\gamma \nabla^\beta Y^{I_3} 
= \frac{1}{2} \left(f(k_2)+f(k_3)-f(k_1)-16\right) \\
\hspace*{7.5cm}
\cdot h(k_1,k_2,k_3)\langle T^{I_1} C^{I_2} C^{I_3} \rangle, \\
\int_{S^4}  Y^{I_1}_{(\alpha\beta)} 
\nabla^{(\gamma}\nabla^{\alpha)}  Y^{I_2} \nabla_\gamma\nabla^\beta  Y^{I_3} 
= \frac{1}{2} \left(\frac{1}{2}f(k_2)+f(k_3)-f(k_1)-16\right) \\
\hspace*{7.5cm}
\cdot h(k_1,k_2,k_3)\langle T^{I_1} C^{I_2} C^{I_3} \rangle, \\
\int_{S^4} \nabla^\gamma Y^{I_1}_{(\alpha\beta)} 
\nabla^{(\alpha}\nabla^{\beta)} Y^{I_2} \nabla_\gamma Y^{I_3} 
= \frac{1}{2} \left(f(k_2)-f(k_1)-f(k_3)-24\right)  \\
\hspace*{7.5cm} \cdot h(k_1,k_2,k_3)
\langle T^{I_1} C^{I_2} C^{I_3} \rangle, \\
\int_{S^4} \nabla_\gamma Y^{I_1}_{(\alpha\beta)} 
\nabla^{(\gamma}\nabla^{\alpha)}  Y^{I_2} \nabla^\beta Y^{I_3} 
= \frac{1}{2} \left(f(k_1)+f(k_2)-f(k_3)-8\right) \\
\hspace*{7.5cm}
\cdot h(k_1,k_2,k_3)\langle T^{I_1} C^{I_2} C^{I_3} \rangle.\\
\end{gathered}
\end{equation}

\renewcommand{\baselinestretch}{1.0} \normalsize

%\nocite{*} %show EVERYTHING in bibliography database (otherwise, get 
%only CITED refs)

\bibliography{strings,m-theory,susy,largeN}
\bibliographystyle{utphys}

\end{document}